\documentclass[10pt,twocolumn,superscriptaddress,aps,prl,balancelastpage]{revtex4-1}
\usepackage{amsfonts,amsmath,amssymb,graphicx,epstopdf,verbatim,dsfont,xcolor}
\usepackage[english]{babel}
\usepackage{subfigure}
\usepackage{epstopdf}
\usepackage{hyperref}
\usepackage{physics}
\hypersetup{
  colorlinks = true,
  citecolor = blue
}
\usepackage{fontawesome}
\usepackage{physics}
\usepackage[all]{hypcap}

\usepackage{cleveref}

\newcommand{\be}{\begin{equation}}
\newcommand{\ee}{\end{equation}}

\graphicspath{{figs/}}
 
\usepackage{amsthm}
\usepackage{bm}

\newtheorem{corollary}{Corollary}
 
\begin{document}

\title{Neural-Network Decoders for Measurement Induced Phase Transitions}

\author{Hossein Dehghani}
\email{hdehghan@umd.edu}
\affiliation{Joint Quantum Institute, NIST/University of Maryland, College Park, MD 20742, USA}
\affiliation{Joint Center for Quantum Information and Computer Science, NIST/University of Maryland, College Park, MD 20742, USA}

\author{Ali Lavasani} 
\affiliation{Joint Quantum Institute, NIST/University of Maryland, College Park, MD 20742, USA}
\affiliation{Condensed Matter Theory Center, University of Maryland, College Park, MD 20742, USA}

\author{Mohammad Hafezi}
\affiliation{Joint Quantum Institute, NIST/University of Maryland, College Park, MD 20742, USA}
\affiliation{Joint Center for Quantum Information and Computer Science, NIST/University of Maryland, College Park, MD 20742, USA}

\author{Michael J. Gullans}
\affiliation{Joint Center for Quantum Information and Computer Science, NIST/University of Maryland, College Park, MD 20742, USA}

\date{today}

\begin{abstract}
Open quantum systems have been shown to host a plethora of exotic dynamical phases. Measurement-induced entanglement phase transitions in  monitored quantum systems are a striking example of this phenomena.  However, naive realizations of such phase transitions requires an exponential number of repetitions of the experiment which is practically unfeasible on large systems. Recently, it has been proposed that these phase transitions can be probed locally via entangling reference qubits and studying their purification dynamics. In this work, we leverage modern machine learning tools to devise a neural network decoder to determine the state of the reference qubits conditioned on the measurement outcomes. We show that the entanglement phase transition manifests itself as a stark change in the learnability of the decoder function. We study the complexity and scalability of this approach in both Clifford and Haar random circuits and discuss how it can be utilized to detect entanglement phase transitions in generic experiments. 
\end{abstract}

\maketitle
\section{Introduction}

Entanglement entropy in closed quantum systems that thermalize generally tends to increase until reaching a volume-law behavior with  entanglement spread throughout the system  \cite{Huse2013Ballistic, Nandkishore2015Many}.
Coupling to a bath profoundly changes the internal evolution of the system \cite{breuer2002theory}, 
which in turn can 
suppress the growth of entanglement and correlations within the system to an area-law behavior \cite{Nayak2013Area, Abanin2013Local}.
A prominent example of such systems  are random quantum circuits with intermediate measurements \cite{Skinner2019Measurement,li2018quantum,Fisher2019Measurement,noel2021observation,Koh22}. In these circuits, where the  unitary time evolution of the system is interspersed by  quantum measurements, the competition between unitary and non-unitary elements leads to a measurement-induced phase transition (MIPT) between a pure phase with an area-law and a mixed phase with a volume-law entanglement behavior \cite{Gullans2020Dynamical, choi2020quantum,jian2020measurement,bao2020theory,zabalo2020critical,Tang2020Measurement,Fuji2020Measurement, Turkeshi2020Measurement, Ippoliti2021Entanglement, Lavasani2021Measurement, Hsieh2021Measurement, Mathias2021Entanglement, Buchhold2021Effective, Altman2021Symmetry, Jian2021Measurement,Czischek21, potter2021entanglement,turkeshi2021measurement,  Altman2022Measurement, Minato2022Fate,Muller2022Measurement, van2022monitoring, koh2022experimental}. 
Such entanglement phase transitions are only accessible when the density matrix is conditioned on the measurement outcomes while they are hidden from any observable which can be expressed as a linear function of the density matrix. On the other hand, to experimentally probe observables which are non-linear functions of the density matrix, one naively needs to reproduce multiple copies of the same state. However, due to intrinsic randomness in measurement outcomes, this naive approach requires repeating the experiment exponentially many times (in system size) \cite{Koh22,Czischek21}.

 Building on the close connection between measurement-induced entanglement phase transitions and quantum error correction \cite{Gullans2020Dynamical,choi2020quantum, Gullans2021Quantum, fan2021self, Li2021Statistical, yoshida2021decoding}, a possible workaround to this obstacle was found in Ref.~\cite{Gullans2020Scalable} for purification transitions, which generically coincide with area-to-volume-law entanglement transitions in random circuit models without symmetry or topological order \cite{potter2021entanglement}. It was shown how to probe these phase transitions through purification dynamics of an ancilla reference qubit that is initially entangled to local system degrees of freedom. 
Subsequently,  the time dependence of the entanglement entropy of the reference qubits signifies the phase transition properties \cite{Gullans2020Scalable, Gullans2020Dynamical,zabalo2020critical}.
To employ this method, one needs to find the density matrix of reference qubits conditioned on the measurement outcomes of the circuit. Hence, the final objective in this approach is to obtain a ``decoder'' that maps the measurement outcomes to the density matrix of the reference qubit. However, such decoders are  only known and implemented for special classes of circuits such as stabilizer  circuits \cite{noel2021observation}. For more generic circuits like Haar-random circuits, finding an analytical solution to this problem is likely unfeasible.

Here, motivated by the  recent successful applications of machine learning algorithms in quantum sciences \cite{carrasquilla2020machine} and especially optimizing quantum error correction codes and quantum decoders \cite{Torlai2017Neural, krastanov2017deep, baireuther2018machine, andreasson2019quantum, nautrup2019optimizing, Liu2019Neural,Siddiqi2020Using,  sweke2020reinforcement}, we provide a generic neural network (NN) approach that can efficiently find the aforementioned decoders. First, we sketch our physically motivated NN architecture. Although we use numerical simulations of Clifford circuits to show the efficacy of our NN decoder, we argue that in principle the same decoder with slight modifications should work for any generic circuit. We investigate the complexity of our learning task by studying the number of circuit runs required for training the neural network decoder.
Importantly, we show that the learning task only needs measurement outcomes inside a rectangle encompassing the statistical light-cone \cite{Gullans2020Scalable,Ippoliti2021Entanglement} of the reference qubit. Furthermore, we demonstrate that by studying the temporal behavior of the learnability of the quantum trajectories, one can estimate the critical properties of the phase transition. Finally, we verify that 
for large circuits one can train the NN over smaller circuits which proves the scalability of our method. 
 
\begin{figure}[]
\centering
\includegraphics[width=8.5cm]{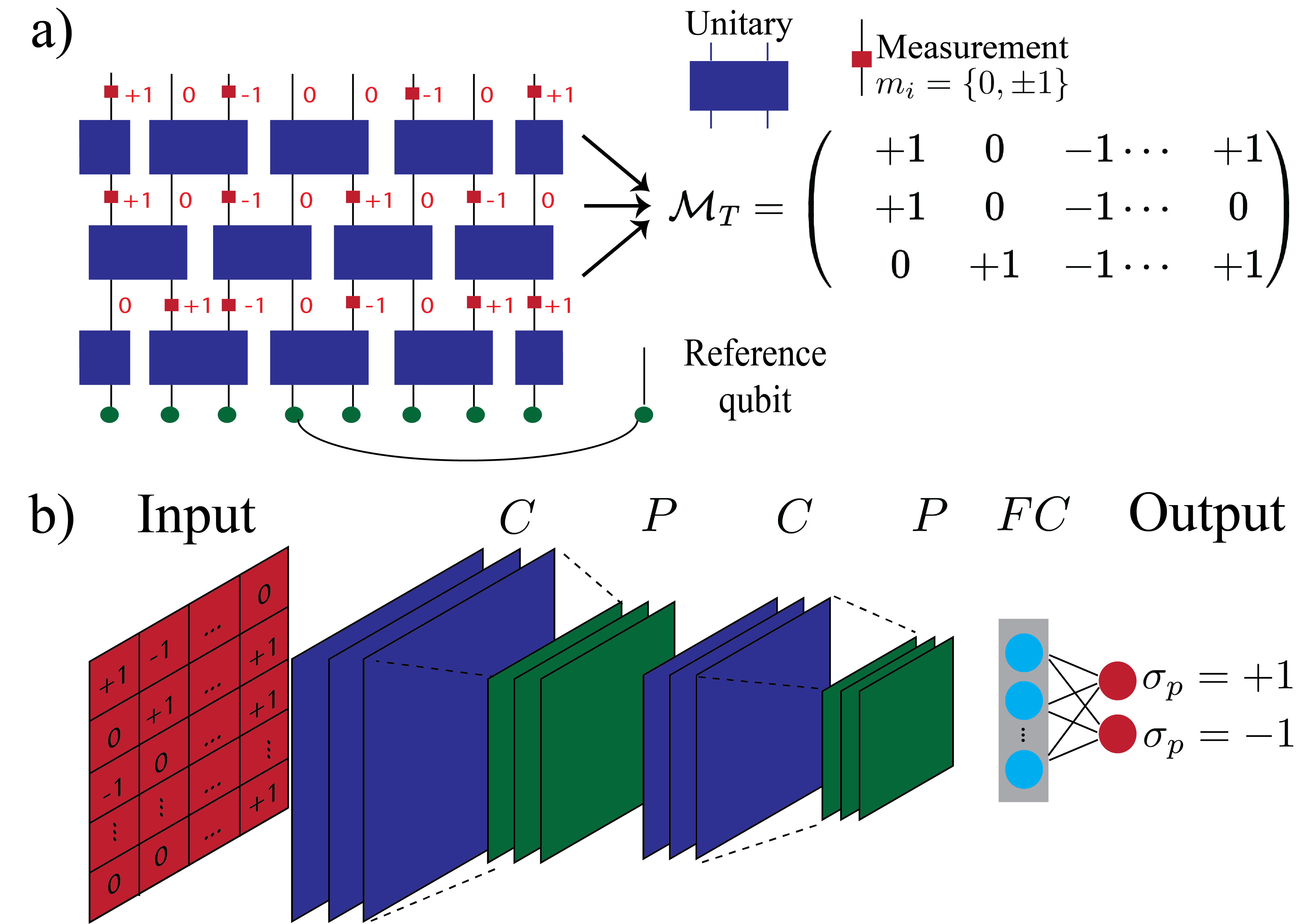}
\caption{(a) Brickwall structure of hybrid circuit with random  two-qubit Clifford gates interspersed with projective $Z$ measurements and with periodic boundary conditions. $\mathcal{M}_T$ denotes the measurement outcome matrix with matrix elements $m_i=\{0 ,\pm 1\}$ ($m_i=0$ when the corresponding qubit is not measured, and $m_i=\pm 1$ when a qubit's Pauli $Z$ is measured).  Here, $T=3$ for this example. (b) Neural network architecture: We use convolutional neural networks composed of (C:convolutional), (P:pooling) and (F:fully connected) layers, trained on quantum trajectories.  The neural network implements a decoder function that predicts the measurement result for the reference qubit $\sigma_p$ using the measurement record in the circuit $\mathcal{M}_T$ as input. } \label{Fig:F1}
\end{figure}
 
\section{Model}
The circuits that we study have a brickwork structure as in Fig.~\ref{Fig:F1}, with $L$ qubits. We consider time evolution with $T$ time steps with repetitive layers of two-qubit random unitary gates, followed by a round of single-site measurements of the Pauli $Z$ operators at each site with probability $p$. 
As one tunes $p$ past some critical value $p_c$, there is a phase transition from a volume-law entanglement behavior ($p<p_c$) to an area-law behavior ($p>p_c$) and a logarithmic scaling at the critical point ($p= p_c$).  Crucially for this work, this phase transition is also manifested in the time dependence of the entanglement entropy of a reference qubit entangled with the system $S_{Q}(t)$ \cite{Gullans2020Scalable}.  $S_{Q}(T)$, averaged over many circuit runs, is known as the coherent quantum information and plays a crucial role in the fundamental theory of quantum error correction \cite{Schumacher96}. For polynomial in system size circuit depths,  $S_Q(T)$ maintains a finite value in the volume-law phase and vanishes in the area-law phase.  The protocol we use to probe $S_Q(t)$ is illustrated in Fig.~\ref{Fig:F1}a. Starting from a pure product state, we make a Bell pair out of the qubit in the middle and an ancilla reference qubit. 
Throughout the paper, we use periodic boundary conditions for the circuit.

\textbf{Decoder}.~To find $S_Q(T)$ in experiment, we need to find the density matrix of the reference qubit at time $T$, which is a vector inside the Bloch sphere and can be specified by its three components $\expval{\sigma_X}$, $\expval{\sigma_Y}$ and $\expval{\sigma_Z}$. Therefore, probing the phase transition can be viewed as the task of finding a decoder function $F_\mathcal{C}$ for a given circuit $\mathcal{C}$, such that
\begin{eqnarray}
F_\mathcal{C}(\mathcal{M}_T)=(\langle \sigma_X \rangle,\langle \sigma_Y \rangle,\langle \sigma_Z \rangle)
\label{decoderMapping}
\end{eqnarray}
where $\mathcal{M}_T$ is the set of circuit measurement outcomes. Let $p_P(m|\mathcal{M}_T)$ for $P\in\{X,Y,Z\}$ denote the probability of getting reference qubit outcome $m=\pm 1$ when measuring $\sigma_P$ of the reference qubit after time $t=T$, conditioned on the measurement outcomes $\mathcal{M}_T$. Since $\langle \sigma_P\rangle=\sum_{m=\pm 1}m~ p_P(m|\mathcal{M}_T)$, the problem of finding the decoder $F_\mathcal{C}$ is equivalent to finding the probability distributions $p_P(m|\mathcal{M}_T)$ for $P\in\{X,Y,Z\}$.

\section{Deep Learning Algorithm} 

Instead of finding $p_P(m|\mathcal{M}_T)$ analytically for a given circuit $\mathcal{C}$, we plan to use ML methods to learn these functions from a set of sampled data points which in principle could be obtained from experiments. The task of learning conditional probability distributions is known as the probabilistic classification task in ML literature \cite{Niculescu2005Predicting, Chuan2017OnCalibration}. Let us fix the circuit $\mathcal{C}$ and the Pauli $P$. A sample data point is a pair of $(\mathcal{M}_T,m)$ for a single run of the circuit where $\mathcal{M}_T$ is the circuit measurement outcomes and $m$ is the of outcome of measuring the reference qubit in the $\sigma_P$ basis at the end of the circuit. By repeating the experiment $N_t$ times, we can generate a training set of $N_t$ data points. By training a neural network using this data set, we obtain a neural network representation of the function $p_P(m|\mathcal{M}_T)$.

Framing the problem as a probabilistic classification task does not necessarily mean that the learning task would be efficient. Indeed, given that the number of different possible $\mathcal{M}_T$ outcomes scales exponentially with the system size, one would naively expect that the minimum required $N_t$ should also scale exponentially for the learning task to succeed, i.e., we need to run the circuit exponential number of times to generate the required training data set. However, the crucial point made in Ref.~\cite{Gullans2020Scalable} is that, when the reference qubit is initially entangled locally to the system, its density matrix at the end of the circuit only depends on the measurement outcomes that lie inside a statistical light cone, and up to a depth bounded by the correlation time that is finite in the system size away from the critical point. Hence, for a typical circuit away from the critical point, the function $p_P(m|\mathcal{M}_T)$ depends only on a finite number of elements in $\mathcal{M}_T$ and that  makes the learning task feasible.  

To show the effectiveness of this method, we test our decoder using data points gathered from numerical simulation of Clifford circuits with $p_c=0.160(1)$ \cite{Fisher2019Measurement}, which enables us to study circuits of large enough sizes. Due to Clifford dynamics, the reference qubit either remains completely mixed at $t=T$ or it is purified along one of the Pauli axis. This means the measurement outcome of $\sigma_P$ at the end of the circuit is either deterministic or completely random. Therefore, it is more natural to view the problem as a hard classification task (rather than probabilistic) where we train the neural network to determine the measurement outcome of $\sigma_P$ (See the methods section). Note, if the reference qubit is purified at the end of the circuit, then the decoder can in principle learn the  decoding function while, if it is not, then the measurement outcomes are completely random, leading to an inevitable failure of the  hard classification. 
Thus, the purification phase transition shows itself as a learnability phase transition. 
It is worth noting that we are only changing how we interpret the output of the NN, i.e. we pick the label with highest probability, so the same NN architecture can be used for more generic gate sets.  
For simplicity, we also only look at the data points corresponding to the basis $P$ in which the reference qubit is purified. 
In an experiment, the purification axis is not known, so one needs to train the NN for each of the three choices of $P$; if the learning task fails for all of them, it means the qubit is totally mixed. Otherwise, the learning task will succeed for one axis and fail for the other two \footnote{Note,  for a fixed Clifford circuit,  the purification axis does not depend on $\mathcal{M}_T$.}, which means the reference qubit is purified. 

Since locality plays an important role in purification dynamics, we employ a particular deep learning \cite{hinton2006reducing, lecun2015deep, goodfellow2016deep} architecture called convolutional neural networks (CNN) that are efficient in detecting local features in image recognition applications \cite{lawrence1997face}. In utilizing  these networks the input data is treated as a snapshot as in Fig.~\ref{Fig:F1}(b) with each pixel treated as a feature of the NN and the label of each image is the measurement outcome of $\sigma_P$.

\section{Learning Complexity}
For a fixed circuit $\mathcal{C}$, we start the training procedure by training the NN with a given number of labeled quantum trajectory measurements,  
and then evaluate its performance on predicting the labels of new randomly generated trajectories produced by the same circuit $\mathcal{C}$. The learning accuracy $1-\epsilon_{l}$ is the probability that the NN predicts the right label. The minimum number of training samples denoted by $M(\epsilon_l)$ to reach a specified learning error $\epsilon_{l}$ can provide an empirical measure of the learning complexity of the decoder function $F_{\mathcal{C}}$ \cite{Bairey2020Learning}. In what follows, we fix the learning error of each  circuit to be $\epsilon_{l} = 0.02$. 

\begin{figure}[]
\centering
\includegraphics[width=8.5cm]{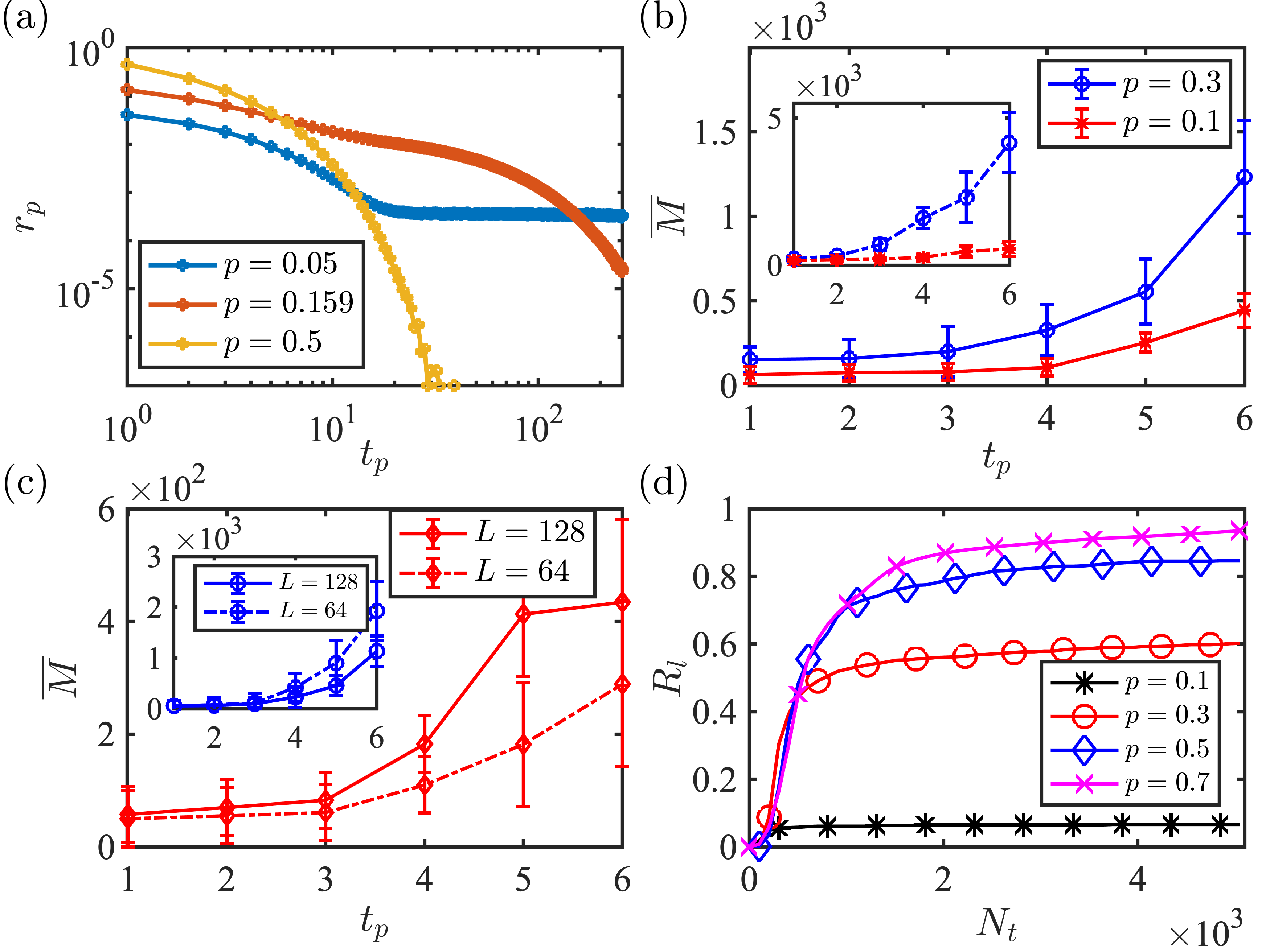}
\caption{(a) Distribution of purified circuits as a function of the purification time for different measurement rates $p=0.05$ (mixed phase), $p= p_c\simeq 0.16$ (critical value), $p=0.5$ (pure phase) with $L=16$ qubits and $N_c=10^7$ random circuits. (b) and (c) Averaged number of quantum trajectories required for learning the reference qubit after conditioning on the purification time $t_{p}$, for $p=0.1$ (mixed phase) and $p=0.3$ (pure phase). Averaging is performed over $N_c=20$ circuits for each $t_{p}$ and error bars are set according to the standard deviation.  In (b) we have circuits with $L=128$ qubits. In the main plot measurement outcomes from inside the fixed light-cone box are used for training while for the inset we use the measurement outcomes from the whole circuit. In (c) we have circuits with $L=128$ qubits (solid-line) and $L=64$ qubits (dashed-line) with $p=0.1$ in the main plot and $p=0.3$ in the inset. (d) Ratio of learned circuits as a function of number of quantum trajectories with $L=64$ and for different $p$ without conditioning on the purification time with $N_c = 10^3$ circuits for each $p$. \label{Fig:F2}}
\end{figure}

In performing this analysis, different learning settings can be considered. Intuitively, for a fixed circuit, we expect the purification time of the reference qubit, $t_{p}$, after which the reference qubit's state does not alter any further, to play an important role in determining $M$. 
Therefore, in our first learning setup, we consider a \textit{conditional} learning scheme where for a given measurement rate, we select quantum circuits based on their purification time $t_p$, which allows us to study the effect of the system size on the learning complexity. Moreover, we discard measurement outcomes corresponding to measurements performed after $t_p$. This is to say that for each $t_p$, measurement outcomes outside a mask with width $L$ and height $t_p$ will be masked. Here, we note that given $N_c$ circuits with the same purification time, in addition to the learning efficiency of each circuit, we need to fix the learning inaccuracy  averaged over $N_c$ circuits, $\delta_l$, which we fix to be $\delta_l=20\%$. We remark that this number is larger than $\epsilon_l$ since some of the conditionally selected circuits have not been learned. 

In the second setting, we remove the conditioning constraint and only consider the overall complexity of the learning task when we randomly generate circuits for a given $p$ in a completely \textit{unconditional} manner. The two schemes can be related using the probability distribution $r_p$ of the purification time as shown in Fig.~\ref{Fig:F2}(a) and explained more concretely in the methods section. 
We should emphasize that the conditional learning scheme is only a tool for studying the complexity of the learning problem for Clifford circuits. For probing the phases and phase transitions in both Clifford and Haar circuits, we use the unconditional learning scheme.
Note that since the reference qubit is entangled locally at the beginning, there is always a finite probability that it will be purified in early times. In the mixed phase, the distribution has an exponentially small tail until exponentially long times (both in system size) whereas in the pure phase, the ancilla purifies in a constant time independent of system size. 
% Since, in principle, all purified trajectories should be learnable by an ideal decoder, this scheme provides an estimate for the purification ratio of the circuits.  
Inspired by the approximate locality structure of hybrid circuits~\cite{Gullans2020Scalable}, we also consider a light-cone learning scheme, where we train the NN using only the measurement outcomes inside a box centered at the middle (see Fig.~\ref{Fig:F4}(b)). 
In Fig.~\ref{Fig:F2}(b), we compare the complexity of the conditional learning task in the pure and mixed phases both by using the light-cone box (main) and whole circuit (inset) measurement data. For each purification time and $p$, we consider $N_c=20$ different circuits and we average over their minimum required training numbers to calculate $\bar{M}(\epsilon_l,\delta_l)$, and show the standard deviation as the error bar. Here, for all the curves, we observe an approximate exponential growth of $\bar{M}(\epsilon_l,\delta_l)$  as a function of the purification time $t_{p}$. By comparing the mixed and pure phases, we notice that the conditional learning task is more complicated in the pure phase than the mixed phase, which is expected since, all else being equal, there are more measurements in the pure phase. Additionally, as shown in the inset, we find that learning with light-cone data is less complicated than using all the measurement outcomes. These behaviors can be understood by recognizing that to learn the decoder we need to explore the domain of the mapping in Eq.\eqref{decoderMapping} whose size scales exponentially with $2^{pTL}$. 

In Fig.~\ref{Fig:F2}(c) we compare the system size dependence of the complexity in the two phases with $L=64, 128$ where we train our networks with the light-cone data. We note that since the size of the light-cone box for a fixed $t_{p}$ is independent of the system-size, we expect the 
%in principle we should expect that
asymptotic complexity to be independent system size. Our numerical observation is partially in agreement with this theoretical expectation.   
In the methods section, this point has been studied further where we explicitly depict the system size dependence of the complexity for circuits with experimentally relevant system sizes $L=\{16, 32, 64, 128\}$.
%because by including the error bars of the two curves we observe that the two curves overlap.\cite{Gullans2020Scalable}. 
In the methods section, we also obtain similar complexity results for circuits with initial states scrambled by a high-depth random Clifford circuit.

In the final step, we consider the unconditional learning task. Fig.~\ref{Fig:F2}(d) shows the ratio of circuits that can be learned, denoted by $R_{l}$, as a function of $N_{t}$, with the circuit depth fixed at $T=10$.

After an initial fast growth in $R_{l}$, the learning procedure slows down. This can be understood by noting that exponentially more samples are required to learn the decoder for circuits with longer purification time. Moreover, the saturation value for each $p$ is bounded by the ratio of circuits that are purified by time $T$, which can be expressed as
\begin{align}
    R_p(T)=\int_0^T~r_p \dd t
\end{align}
where $r_p$ is the purification rate plotted in Fig.~\ref{Fig:F2}(a).

\begin{figure}[]
\centering
\includegraphics[width=8.5cm]{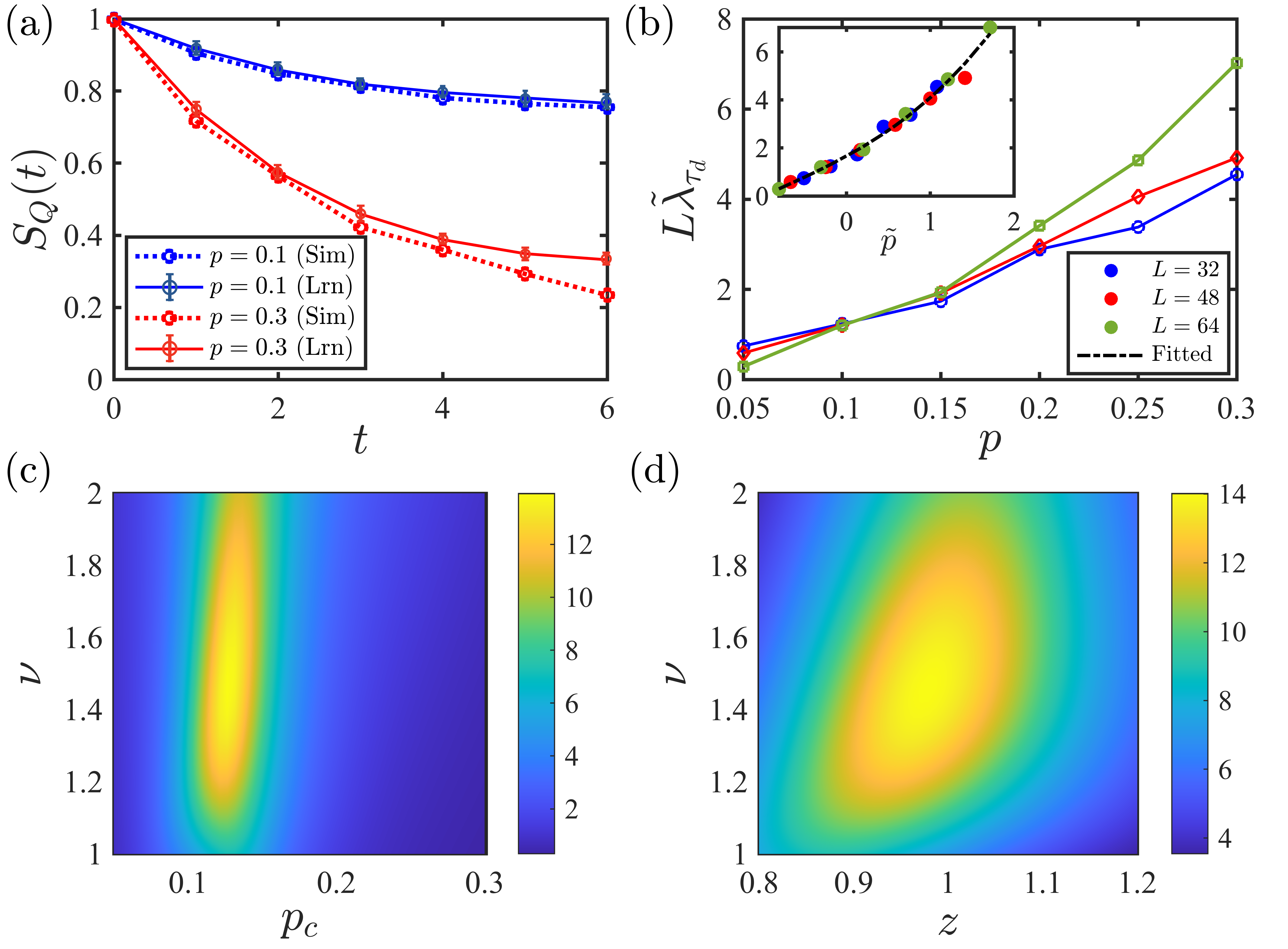}
\caption{Temporal behavior of the reference qubit's entanglement entropy averaged over $N_c=10^3$ circuit configurations for a given $p$. (a) Comparing the temporal behavior for a circuit with $L=64$ qubits in the mixed ($p=0.1$) and pure ($p=0.3$) phases. The dashed and solid lines are achieved from learning quantum trajectories, and exact simulation of the circuits, respectively. Each point in these curves has a statistical error less than $2\%$. (b) Scaled temporal rate of the learned entanglement entropy, $L\tilde{\lambda}_{\tau_d}$, as a function of the measurement rate at a fixed scaled time $\tau_d=t_d/L=1/16$. Inset: Collapsing the curves for $L^z \tilde{\lambda}_{\tau_d}$ as a function of $\tilde{p}=(p-p_c)L^{z/\nu}$, using $p_c=0.13, \nu=1.5,$ and $z=1$. (c)  Inverse fitting error as a function of $\nu$ and $p_c$ for $z=1$. (d) Inverse fitting error as a function of $\nu$ and $z$ for  $p_c=0.13$. In (c) and (d) yellow areas show the best parameter estimates for the phase transition. \label{Fig:F3}}
\end{figure}

\section{Dynamics of Coherent Information}
We can utilize the NN decoder to study the critical properties of  the phase transition. 
For a fixed circuit configuration $c$ with a given $p$, let $\rho_c$ and $s_c(t) = -\mathrm{tr}(\rho_c \log_2 \rho_c)$ denote its density matrix and von Neumann entropy of the reference qubit after time $t$, respectively. Based on this definition, we let $S_Q(t)=-\sum_{c=1}^{N_c} \frac{1}{N_c}{\mathrm{tr}(\rho_c \log_2 \rho_c)}$ denote the average entropy of the reference qubit after time $t$, i.e., the coherent quantum information of the system with 1 encoded qubit.  We  may assume on general grounds that $S_Q(t)$ follows an early time exponential decay $e^{-\lambda t}$ with $\lambda$ following the scaling form
\begin{eqnarray}
\lambda  = L^{-z} f[(p - p_c)L^{z/\nu}], \label{scaling}
\end{eqnarray}
where $z$ and $\nu$ are the dynamical and correlation length critical exponents respectively \cite{noel2021observation}. In stabilizer circuits, the density matrix of the reference qubit will be either purified completely with $s_c=0$, or will be in a totally mixed state with $s_c=1$. 
Since $S_Q(t)$ and the ratio of purified circuits $R_{p}(t)$ are related by $S_{Q}=1-R_{p}$, we can estimate $S_Q(t)$ by the ratio of learnable circuits of depth $t$ in the unconditional scheme described above.  We denote the estimated value of $S_Q(t)$ from learning by $\tilde{S}_{Q}$. More concretely: $(1)$ For each given $p$ and $L$ we generate $N_c=10^3$ random circuits and we evolve them for $T\sim \mathcal{O}(10)$ time steps which does not scale with the system size and record the measurement outcomes $\mathcal{M}_T$, $(2)$ At the end of this time evolution, we measure the spin of the reference qubits along the purification axis, $m$, (3) For each circuit we use the corresponding labeled data $(\mathcal{M}_T,m)$ and we train our neural network with this data to make future predictions. We note that since in this approach there is no constraint in generating the circuits and their quantum trajectories, this procedure can be directly applied to experimental data without requiring any post-selection or conditioning procedure.

In Fig.~\ref{Fig:F3}(a) we compare the temporal behavior of the coherent information obtained from an ideal decoder  and the NN decoder introduced here where for each $p$ we consider $N_c=10^3$ different circuit configurations. As demonstrated in Fig.~\ref{Fig:F3}(a), in the mixed phase the learned entanglement entropy closely follows the simulated entanglement entropy, while in the pure phase the two curves start to deviate from each other after a few time steps. This behavior is consistent with previous observations in Fig.~\ref{Fig:F2} where we demonstrated that the learning task is easier in the mixed phase. 
Since at the critical point this phase transition can be described by a $1+1$-D conformal field theory \cite{Skinner2019Measurement,Fisher2019Measurement}, the dynamical critical exponent can be fixed in advance $z=1$ and correspondingly we define the scaled time $\tau=t/L$. Furthermore, since the argument of the scaling function $f$ on the right hand side of Eq.\eqref{scaling} becomes independent of $L$ at $p_c$, we expect to see a crossing in 
\begin{eqnarray}
L\lambda_{\tau_d} \approx |\frac{d \ln S_Q}{d\tau}|_{\tau_d}. 
\end{eqnarray}
when it is plotted for different system sizes. Here, $\tau_d=t_d/L$ is the differentiation time which should be sufficiently large. In Fig.~\ref{Fig:F3}(b), we evaluate the decay rate obtained by learning, $\tilde{\lambda}_{\tau_d}$, for three different system sizes, $L=\{32,48,64 \}$, at $\tau_d = 1/16$ using $\tilde{S}_{Q}$. The corresponding times are $t_d=\{2, 3, 4\}$ for which the deviation of the learned and simulated coherent information is negligible. Here, we notice an approximate crossing in the region $0.1 \lesssim p_c \lesssim 0.15$ signaling a phase transition in this region. 

More systematically, we may find the best estimated values of the critical data by collapsing the decay rate curves according to the scaling ansatz in Eq.\ref{scaling}. In particular after fixing $z=1$, we can search simultaneously for $p_c$ and $\nu$ so that the fitting error of the regression curve would be minimized (See the methods section). The inverse error has been plotted as a function of $p_c$ and $\nu$ in Fig.~\ref{Fig:F3}(c) where we observe that the lowest error corresponds to the region $p_c\simeq 0.13, \nu \simeq 1.5$. Similarly, we can examine our assumption about the conformal symmetry of the transition, by fixing $p_c=0.13$, and allowing $\nu$ and $z$ to vary as in Fig.~\ref{Fig:F3}(d). Here, we observe that the lowest error corresponds to the region around $\nu\simeq 1.5, z \simeq 1$. Using the obtained estimates, namely, $\nu \simeq 1.5, z\simeq 1,$ and $p_c\simeq 0.13$, in the inset of Fig.~\ref{Fig:F3}(b) we collapse the three curves of $L^z\tilde{\lambda}$ as a function of $\tilde{p} = (p-p_c)L^{z/\nu}$. In the methods section, we search simultaneously over all  three parameters and find that the best estimates for the critical data are in the region $p_c=0.14\pm 0.03$, $z=0.9\pm 0.15$, and $\nu=1.5\pm 0.3$.  These results are in good agreement with the exact results obtained from the half-chain entanglement entropy, $z=1, p_c\simeq0.16,$ and $\nu\simeq 1.3$ \cite{Fisher2019Measurement, Skinner2019Measurement}. Additionally, we verify our learning results by comparing them with the results obtained from exact simulations of $S_{Q}(t)$, where we demonstrate that by increasing $L$ and $N_c$, the phase transition parameters can be determined more accurately.

\section{Scalability of Learning}
\begin{figure}[]
\centering
\includegraphics[width=8.5cm]{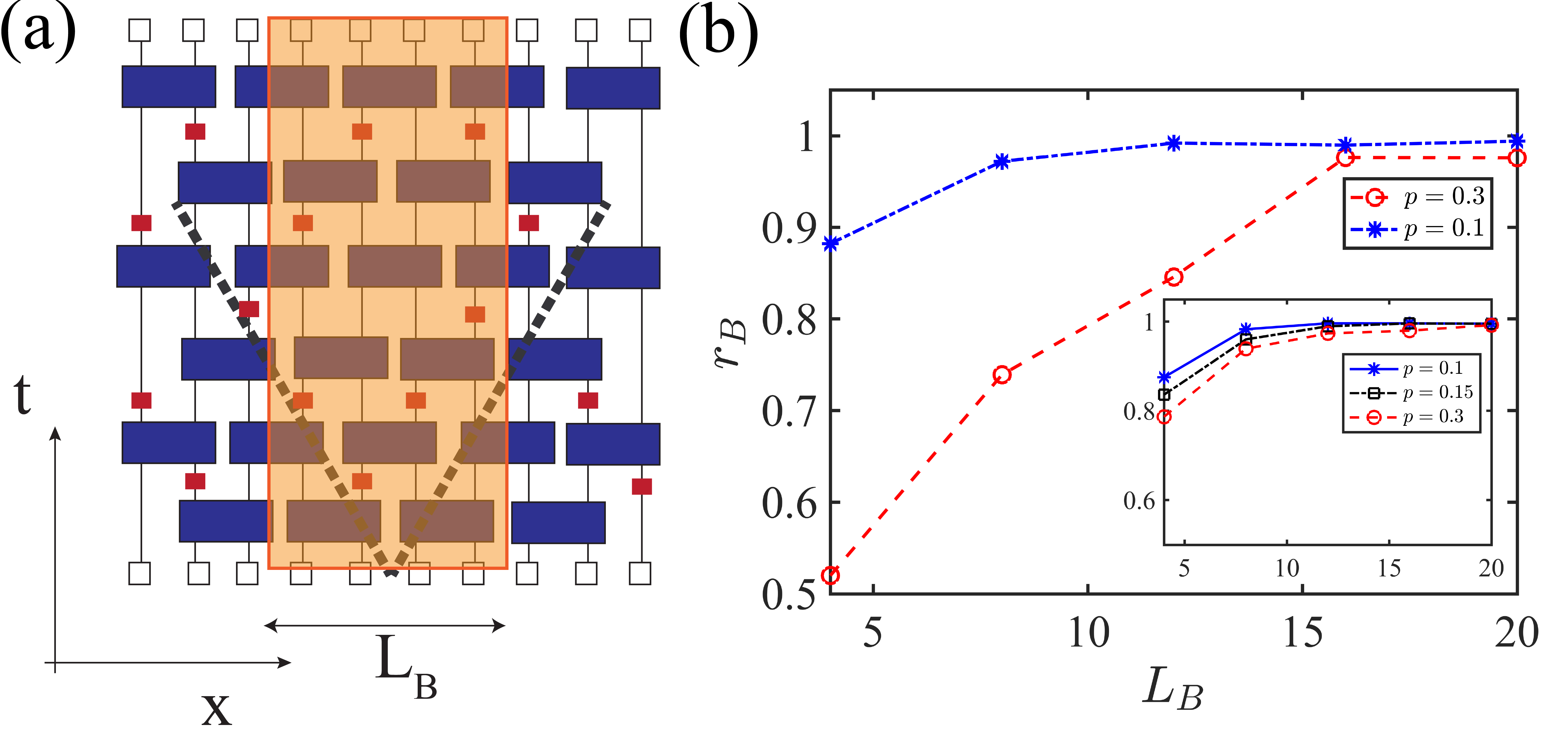}
\caption{(a) Predicting the decoder function of a circuit using the neural network trained by the measurement outcomes inside the small circuit in the orange box. (b) Fraction of experiments that can be learned using smaller circuits of width $L_B$.  When the ratio is 1, that means that there is no benefit in the training from increasing $L_B$. Main: $L=64$ and $N_c=100$. Inset: $L=32$ and $N_c=200$. }\label{Fig:F4}
\end{figure}

An important feature of a practical decoder is the possibility of training it on small circuits and then utilizing it for decoding larger circuits. %\cite{breuckmann2018scalable}. 
Here, due to the approximate locality of the temporal evolution of the random hybrid circuits, one can examine the scalability of the decoders in a concrete manner. For a given circuit with $L$ qubits, we generate smaller circuits with $L_B<L$ number of qubits which have identical gates as the original circuit in a rectangular narrow strip around the middle qubit which is entangled to the reference qubit. The geometry of the two sets of circuits is displayed in Fig.~\ref{Fig:F4}(a) where the depth of the two sets of circuits are chosen to be equal. Here, for each $p$ we generate $N_c$ large circuits with $L=\{32, 64\}$ and $T=10$ time steps. We also only consider those circuits that are learnable using measurement outcomes from the original circuit. Next, for each of these circuits, for $L_B=\{4, 8, \cdots, 20\}$ we generate their corresponding smaller circuits and we run them to generate $N_{t}=5\times 10^3$ quantum trajectories. In the training step, we use the quantum trajectories produced from the smaller circuits to train our neural networks. In the testing step, however, we use these neural networks to make prediction for the quantum trajectories obtained from the larger circuits. As we observe in Fig.~\ref{Fig:F4}(b), by increasing $L_B$ the ratio of the circuits that can be learned by the smaller circuits' NNs increases. Also, consistent with the effective light-cone picture, we see that for both system sizes, $L=\{32, 64\}$, the largest required $L_B$ to reach almost a full efficiency, according to the light cone condition can be determined by $L_B\gtrsim 2T$ which in our case corresponds to $L_B=20$. This demonstrates that independent of the system size, the light-cone-trained NNs can be used for learning larger circuits.

\section{Generalization to Haar Random Circuits}

To benchmark the methods, we have focused on Clifford circuits, which have two important simplifications for our learning procedure.  First, the purification axis is independent of the measurement outcomes and the learning only needs to be performed along one of the $\{X, Y, Z\}$ axes in the Bloch sphere. In addition, the purification occurs at specific layer of the circuit.  Therefore, it is important to test our results in more generic Haar random circuits, where the purification axis can be along any radius in the Bloch sphere and purification dynamics occurs throughout the circuit evolution \cite{zabalo2020critical}. Here, we show how to adapt our method to Haar random circuits to see clear evidence of the two phases. We leave the study of critical properties of the entanglement phase transition with our method for future work. 

\begin{figure}
    \centering
    \includegraphics[width=7cm]{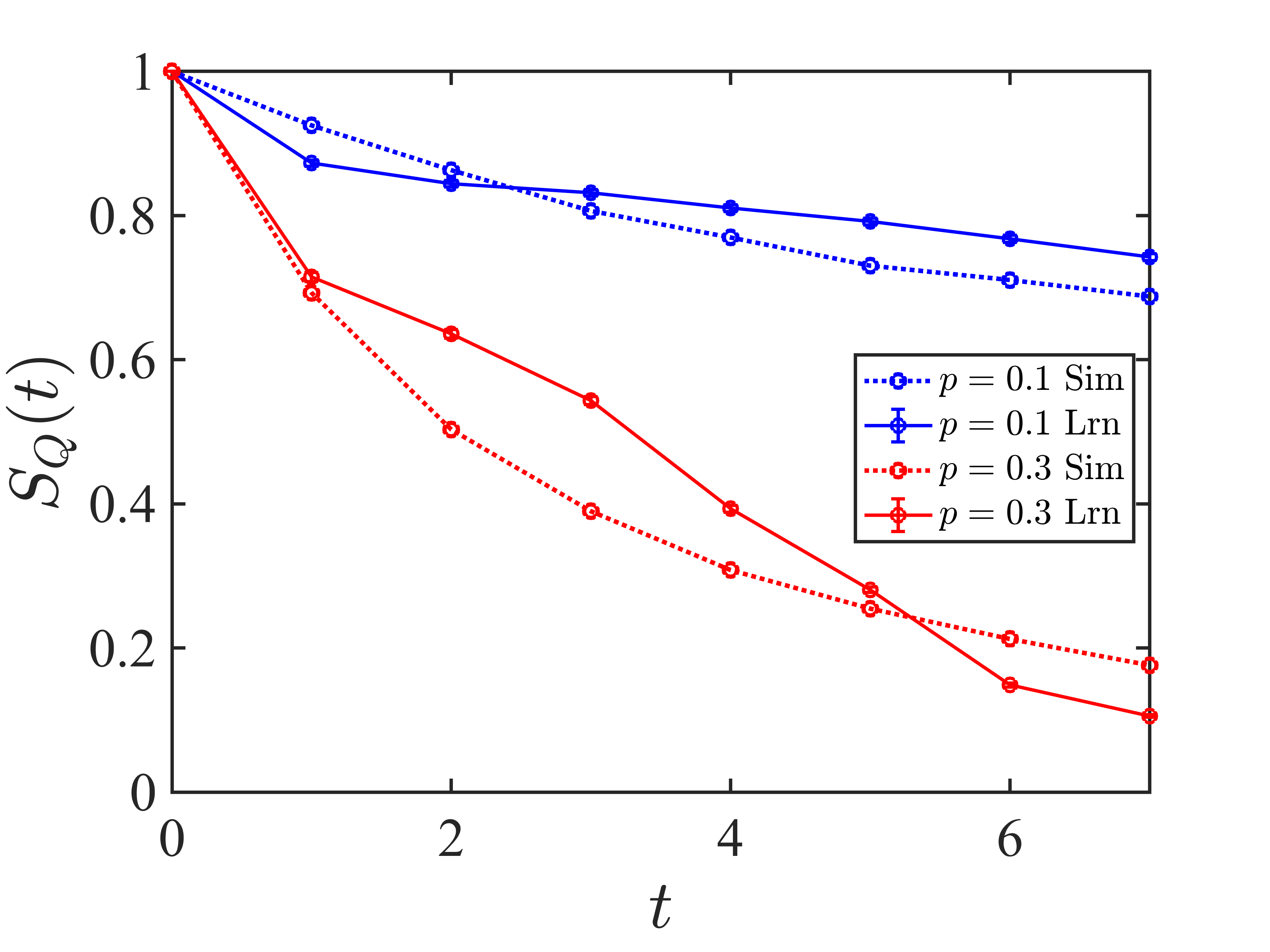}
    \caption{Simulated and learned $S_Q(t)$ obtained for Haar random  circuits.  We took a system of size $L=8$ with 100 random circuit realizations.  Training was performed on 5000 trajectories per circuit. Statistical errors based on the variance of the mean $S_Q(t)$ are less than $1\%$.}
    \label{fig:Fig5}
\end{figure}

To obtain the decoder function $F_\mathcal{C}$ for generic circuits, we need to create three independent sets of labeled data for measuring $\sigma_{i}$ with $i \in \{X,Y,Z\}$ obtained from quantum trajectories. Next, these three sets of labeled measurement data, represented by $\{\mathcal{M}^{i}_T, m_i\}$, are used to train three independent neural networks to produce the probability distribution of reference qubit density matrix expectation values $p_i(m|\mathcal{M}_T)$. Consequently, given new quantum trajectories, the trained $p_i$'s will be employed to estimate $\langle \sigma_i\rangle$. Finally, using standard density matrix tomography methods, such as the maximum likelihood estimation of the density matrix of a single qubit \cite{James2001Measurement}, we can obtain the most likely physical density matrix associated with the predicted $\langle \sigma_i\rangle$'s. An illustrative example of the learning dynamics in the two phases for a small number of circuits is shown in Fig.\ref{fig:Fig5} where we study $S_Q(t)$ and its learned value as a function of time for a circuit with $L=8$ qubits in the two phases ($p_c \approx 0.17$ for this model \cite{zabalo2020critical}). We see from this example that our NN decoder straightforwardly generalizes to generic quantum circuits and using a larger circuit ensemble and quantum trajectories it should be possible to study the phase transition properties. 

\section{Discussion and Outlook}
%\hd{This can be removed: In this work we designed a neural network decoder that predict the state of the reference qubits which can locally measurement induced phase transitions in hybrid random Clifford circuits. }

As a main future direction to explore, we note that from an experimental perspective, it is possible incorporate different errors, which are common in the realization of the two-qubit gates and/or measurement processes, in our machine learning framework. An intriguing possibility is to find neural network decoders that are successful in learning deep circuits with local data  \cite{yoshida2021decoding}. Similarly, implementing neural network decoders for other MIPTs such as systems with long-range interactions \cite{Minato2022Fate}, and symmetric MIPT \cite{barratt2022transitions}, is an immediate extension of this work. Another intriguing question is to investigate whether it is possible to use our decoder approach for MIPTs where it is not equivalent to purification transitions. In the context of quantum error correction and fault-tolerance, the purification dynamics in measurement-induced phase transitions leads to a rich set of examples of dynamically generated quantum error correcting codes \cite{Brown13,Gullans2020Dynamical,Gullans2021Quantum,Hastings21}. Designing similar decoders as considered here for other types of dynamically generated logical qubits is a rich avenue of investigation. We also highlight that our empirical complexity results raise interesting questions about the complexity of learning an effective Hamiltonian description \cite{Anshu21, Haah21, van2022monitoring} of the measurement outcome distributions for monitored quantum systems. Finally, we note that improving our neural network algorithms to find the optimal decoder, and investigating the applicability of unsupervised machine learning techniques for this problem is left for future studies \cite{Huang2022Predicting, Dehghani2022Unsupervised}.

\section{Code Availability}
The code used in this study is available from the corresponding author upon  request. 

\begin{acknowledgments}
\textbf{Acknowledgments}.---We acknowledge stimulating discussions with Alireza Seif, David Huse, Pradeep Niroula, Crystal Noel, Grace Sommers, and Christopher White. We acknowledge support from  the National Science Foundation (JQI-PFC-UMD and QLCI grant OMA-2120757). H.D. and M.H. acknowledge support from
ARO W911NF2010232, AFOSR FA9550-19-1-0399, NSF OMA-2120757, QSA-DOE and Simons and Minta Martin foundations. This work used the Extreme Science and Engineering Discovery Environment (XSEDE), supported by the grant number PHY210049, at the Pittsburgh Supercomputing Center (PSC) \cite{xsede}. M.H. thanks ETH Zurich for their hospitality during the conclusion of this work. 
\end{acknowledgments}

\section{Methods}
\textbf{Quantum Dynamics.}
The dynamics of hybrid circuits considered in this work in general can be described using the quantum channel formalism. The wave function of the circuit, denoted by $|\psi_S\rangle$ at the beginning of time evolution is entangled to a reference qubit. Formally, the time evolution of the system under this setting can be modeled using Kraus operators  \cite{nielsen2002quantum}, 
 \begin{eqnarray}
K_{\vec{m}} = U_t P^{m_t}_t \cdots  U_1 P^{m_1}_1
 \end{eqnarray}
where $m_t$, $U_t$ and $P_t^{m_t}$, denote the measurement outcomes, unitary gates, and projective measurements at the $t$-th layer of the circuit, respectively. We also denote the set of all measurement outcomes in different layers via $\vec{m}$. The corresponding evolution of the density matrix, $\rho$, can be described via the following quantum channel,
\begin{eqnarray}
\mathcal{N}_t(\rho) = \sum_{\vec{m}}K_m \rho K_m^{\dagger} \otimes |\vec{m}\rangle \langle \vec{m}|.
\end{eqnarray}
For our purpose, to generate the quantum trajectories we need to consider the time evolution of the system at the level of the wave functions. Under an arbitrary unitary operator $U$, the wave function evolves as 
\begin{align}
    |\psi\rangle \rightarrow U|\psi\rangle.
\end{align}
For projective measurements, we consider a complete set of orthogonal projectors with eigenvalues labeled by $m$ satisfying $\sum_{m} P^m_t=1$ and $ P^{m}_t P^{m'}_t=\delta_{m m'}P^m_t$ under which the wave function evolves as,
\begin{align}
    |\psi\rangle \rightarrow \frac{P^{m}_t|\psi\rangle }{||P^{m}_t|\psi\rangle ||}.
\end{align}

In simulating the time evolution of the wave functions, we use random unitaries sampled from the Clifford group where, under any conjugation operation, the Pauli group is mapped to itself \cite{Gottesman1998Theory}. Such circuits, according to the Gottesman-Knill theorem, can be classically simulated in polynomial times in the system size \cite{gottesman1998heisenberg, aaronson2004Improved}.

%For Clifford circuits with $L$ qubits, we consider time evolution trajectories with $T$ time steps, where we define each time step to be composed of the application of adjacent random unitary gates followed by $Z$ measurements at random sites. 

\textbf{Implementation of Deep Learning Algorithms.}
In this work we mainly used convolutional neural networks for learning the decoder function. These network are composed of several interconnected convolutional and pooling layers. The convolutional layer uses the locality of the input data to create new features from a linear combination of adjacent features through a convolution process. These layers are followed by pooling layers which reduce the number of features. Finally, a fully connected layer is used to associate a label to the newly generated features, thus classifying the data. These layers can be repeated a number of times for more complicated input data. 
Our neural network architecture symbolically displayed in Fig.~\ref{Fig:F1}(b) consists of eight layers whose hyperparameters are chosen by an empirical parametric search to optimize the learning accuracy when the number of samples are smaller than $5\times 10^4$. From left to right these layers include: (1) a convolutional layer with a $L_q/2$ filters where $L_q$ is the number of qubits with a kernel size of $4 \times 4$, and a stride size of $1 \times 1$ with a rectified linear unit (ReLu) activation function, (2) a convolutional layer with a $L_q/2$ filters where $L_q$ is the number of qubits with a kernel size of $3 \times 3$, and a stride size of $1 \times 1$ with a Relu activation function, (3) a maximum pooling layer with a window size of $2\times 2$ to decrease the dimension of the input data, (4) a dropout layer with a dropping rate of $r_d =0.2$ to prevent overfitting, (5) a flattening layer to convert the data into a one-dimensional vector, (6) a dense fully connected layer with a Relu activation function whose number of output neurons is variable and is determined according to the number of training samples, $N_{n} = 512*(1 + 2 \lfloor N_{t}/ 2000\rfloor )$ where $\lfloor x \rfloor$ denotes the floor function of $x$, (7) a dropout layer with a dropping rate of $r_d =0.2$, (8) a dense fully connected layer with a sigmoid activation function which generates the prediction for the spin of the reference qubit. Finally, since we have a classification problem, the loss function for comparing the predicted labels and the actual labels is a binary cross entropy function. Using this loss function, for training our neural network model, we use the Adam optimization algorithm with a learning rate $l=0.001$. The implementation of our neural network layers and their optimization was done by the Python deep-learning packages TensorFlow and Keras. 
\appendix
\section{Scaling analysis and estimation of critical exponents}
The critical exponents of this measurement induced phase transition can be investigated from the decay rate of the reference qubit's entanglement entropy denoted by $\lambda$, which has the scaling (see Eq.~\ref{scaling})
\begin{eqnarray}
L^{z}\lambda  = f[(p - p_c)L^{z/\nu}] \label{scalingApp}.
\end{eqnarray}
While in the main text we fixed $z=1$ based on the assumption of conformal invariance, here, we perform the analysis with $z$ allowed to vary. To find the best combination of the critical data that collapses our data according to this ansatz, we compare the normalized mean squared errors (NMSE), $\varepsilon_{\textrm{NMSE}}$ such that the best fit is obtained when $\varepsilon_{\textrm{NMSE}}^{-1}$ is maximized \cite{cichosz2014data}. In particular, for a given $p_c, \nu,$ and $z$, using cubic polynomials we first find the regression curve of $y\equiv L^z \lambda$ as a function of $(p-p_c)L^{z/\nu}$, and then we evaluate the corresponding value of the mean squared error between $y$ and the best fitted value of it $\hat{y}$. We point out that 
%since on the left side of this equation we have $L^z \lambda$, which varies for different values of $z$, 
in order to compare mean squared errors for different combinations of $(p_c, \nu, z)$, we have to normalize the data by defining dimensionless deviations and then evaluate the NMSE for different combinations of critical data.

\begin{figure*}
    \centering
    \includegraphics[width=18cm]{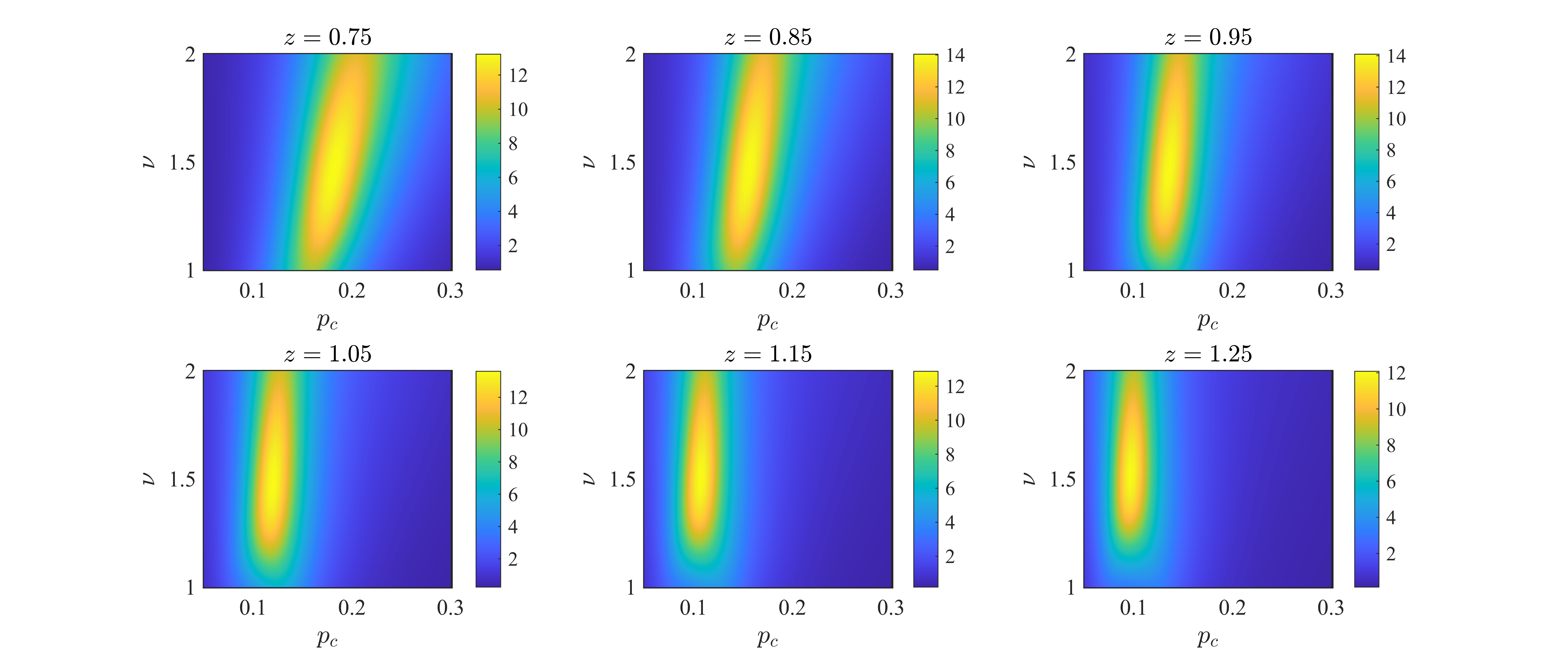}
\centering
    \caption{Finding the best fit for critical properties using maximum inverse normalized mean squared error $\varepsilon^{-1}_{\textrm{NMSE}}$ of the learned decay rate $\tilde{\lambda}$. Intensity of the colormap  $\varepsilon^{-1}_{\textrm{NMSE}}$ plotted as a function of fitted $p_c$ and $\nu$ for different dynamical scaling exponent $z$ displayed on the top of each subplot. The best fits (highest inverse error) are obtained for $z=0.85$, and $z=0.95$. }
    \label{fig:variableZ}
\centering
\end{figure*}

The results of this analysis are displayed in Fig.~\ref{fig:variableZ}, where we have plotted $\varepsilon_{\textrm{NMSE}}^{-1}$ as a function of $\nu$ and $p_c$ for $6$ different values of $z$ ranging from $0.75$ to $1.25$. Based on the subplots in this figure, we observe that the highest values for $\varepsilon_{\textrm{NMSE}}^{-1}$ are obtained for $z\simeq 0.85-0.95$ which is quite close to the value expected from theoretical results based on conformal symmetry $z=1$. Allowing $\varepsilon_{\textrm{NMSE}}^{-1}$ to vary within  almost $10\%$ of its maximum value, we obtain following range for the best fits of the critical data, $p_c=0.14\pm 0.03$, $\nu=1.5\pm 0.3$, and $z=0.9\pm 0.15$.

\begin{figure}
    \centering
    \includegraphics[width=8.5cm]{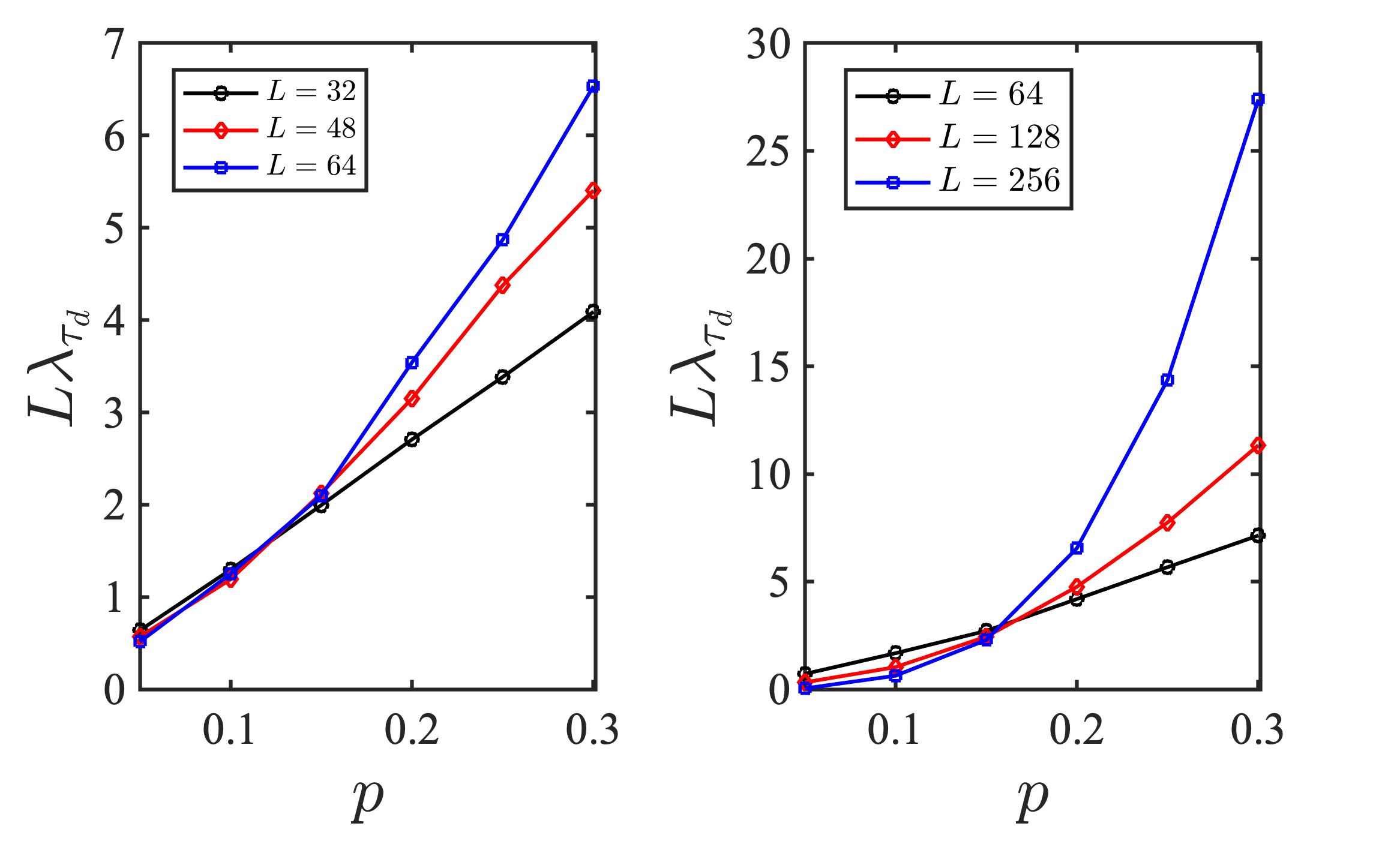}
    \caption{Rate of the simulated entanglement entropy $S_Q$ (not learned) as a function of the measurement rate for different system sizes obtained from exact stabilizer simulation of Clifford circuits for different system sizes. The crossing point of the scaled decay rates, represent the critical measurement rate. (Left) The dimensionless time is $\tau_d=t_d/L=1/16$ and the system sizes are the same as those used for the learning protocol simulations $L=32, 48, 64$. (Right) Scaled decay rates of the entanglement entropy of the reference qubit obtained for larger system sizes $L=64, 128, 256$. }
    \label{fig:lambdaExactSim}
\end{figure}

Finally, we compare our results with the results obtained directly from exact numerical simulations of Clifford circuits without employing our learning scheme. The results of such simulations for the decay rates for different system sizes have been displayed in Fig.~\ref{fig:lambdaExactSim}. In the left subplot we have shown the results for the same system sizes as used for our learning simulations where we observe a crossing of the curves at $p_c\simeq 0.13$ which supports our results obtained from the learning scheme. Furthermore, in the right subplot we observe that for larger system sizes, the obtained crossing of the curves is around $p_c\simeq 0.16$ which is very close to the results obtained from half chain entanglement entropy \cite{Fisher2019Measurement, Skinner2019Measurement}. Accordingly, we expect that by increasing $L$ and $N_c$, the estimates obtained from our learning scheme should improve. 

\appendix
\section{Key measurements in Clifford circuits}
Consider a hybrid Clifford circuit $\mathcal{C}$ which has $M$ Pauli measurements. Imagine applying this circuit on an initial stabilizer state which is entangled to a reference qubit. Assume that as a result of this, the reference qubit disentangles and purifies into the $\ket{P;p_R}$ state, where $P$ is one of the Paulis and $p_R=\pm 1$ determines which eigenvector of $P$ the reference qubit has been purified into. Let $s_1,\cdots, s_M=\pm 1$ denote the measurement outcomes for a single run of the circuit. If we run the same circuit again, the ancilla will purify in the same basis $P$, but we may get different $p_R$ as well as different $s_i$. The goal is to understand the relation between the value of $p_R$ and the measurement outcomes $\{s_i\}_{i=1}^M$. 

When a Pauli string is measured on a stabilizer state, the result is either predetermined (in case the Pauli string is already a member of the stabilizer group up to a phase) or it is $\pm 1$ with equal probability. We call the former determined measurements and the latter undetermined measurements. Note that in a stabilizer circuit, whether a measurement is determined or undetermined is independent of previous measurement \textit{outcomes}. Therefore, for a given circuit $\mathcal{C}$ and a fixed ordering of performing measurements, it is well defined to label measurements as either determined or undetermined without referring to a specific circuit run. 

The following is a straightforward result of the Gottesmann-Knill theorem:
\begin{corollary}\label{crl_pr}
There exists a \textit{unique} subset of undetermined measurement results $\{s_{j_1},\cdots,s_{j_m}\}$ (which we call key measurements) such that,
\begin{align}\label{equ_crl_pr}
    p_R\times s_{j_1}\times s_{j_2}\times \cdots \times s_{j_m}=c
\end{align}
where $c=\pm 1$ is the same for all circuit runs. We call this set the key measurements set. 
\end{corollary}

Note that since key measurements are undetermined measurements, their value are independent of each other. Hence, to predict $p_R$ from undetermined measurement outcomes with any accuracy better than $1/2$, one needs to have access to all key measurement results. 

Each determined measurement can be seen as a constraint between previous undetermined measurement outcomes. Specifically, if $s_i$ is a determined measurement result for some $i$ it means that there is some fixed $c'=\pm 1$ (independent of circuit run) and a subset of  undetermined measurements $\{s_{j'_1},\cdots, s_{j'_m}\}$ such that
\begin{align}
    s_i \times s_{j'_1}\times \cdots \times s_{j'_m}=c'
\end{align}
The similarity to the Corollary \ref{crl_pr} is not accidental: if the reference qubit is purified in the $P$ Pauli basis, it means that measuring it in the $P$ basis would be a determined measurement. 

Existence of these constraints then means that if we relax the condition of the measurements being undetermined in Corollary \ref{crl_pr}, then the set of key measurements is no longer unique; we may be able to replace some measurement outcomes in Eq.\eqref{equ_crl_pr} with a product of others using the constraints between measurement outcomes.

\section{Relation between conditional and unconditional learning schemes}
Here, under certain conditions, we argue that the results of the two learning schemes as displayed in Fig.~\ref{Fig:F2} are related to each other. In particular, using the purification-time distribution of the circuits in Fig.~\ref{Fig:F2}(a), learnability $R_{l}(N_{t})$, is related to the purification ratio $r_{p}(t_{p})$. In what follows to make our analysis more intelligible, we assume that the learning error is nearly vanishing, $\epsilon_l\simeq 0$. Next, we need to study the averaged learning efficiency of our decoder which for a given $t_{p}$ and $N_{t}$ we denote by $\eta_l(t_{p}, N_{t})$. For a given $t_p$ and $N_t$, this quantity is related to the averaged inaccuracy introduced in the text by $\eta_l=1-\delta_l$. To proceed, we employ a simplifying assumption which is approximately consistent with our numerical results. More concretely, we imagine a decoder with a sharp step-like behavior for $\eta_l(t_{p}, N_{t})$ as a function of $N_{t}$. Using the Heaviside theta function $\theta_{H}(x)$, we suppose $\eta_l(t_{p}, N_{t})=\theta_H(N_t-M(t_p))$ where ${M}(t_p)$ is the minimum number of training samples to reach full efficiency for $t\leq t_p $. From the definitions, if follows straightforwardly that
\begin{eqnarray}
R_{l}(N_{t}) = \sum_{t_{p}=1}^{t_{p}^{\mathrm{Max}}(N_{t})} r_{p}(t_{p}), \label{R_l} 
\end{eqnarray}
where $t_{p}^{\mathrm{Max}}(N_{t})$ is the maximum purification time that can be learned for a given $N_t$. However, this quantity can be evaluated by inverting the function $M(t)$ according to $t_{p}^{\mathrm{Max}}(N_{t})=M^{-1}(N_t)$ where $M^{-1}(N_t)$ is the inverse function of $M(t_p)$. Now, we notice that $M(t_p)$ after averaging over different circuits, can be read from the averaged minimum number of training samples in Fig.~\ref{Fig:F2}(b). Therefore, by integrating the information in Fig.~\ref{Fig:F2}(a) and Fig.~\ref{Fig:F2}(b) plus $\eta_l(t_{p}, N_{t})$, one can explain the behavior of $R_{l}(N_{t})$ in Fig.~\ref{Fig:F2}(d). Here, although we do not have the explicit form of  $\eta_l(t_{p}, N_{t})$, we use the step-like behavior as an approximation which is justifiable due to the exponential behaviors of the complexity as a function of the purification time. Thus, using Eq.\ref{R_l} as a plausible approximation for the learnability of our decoder, we expect that during the initial fast growth of the curves in Fig.~\ref{Fig:F2}(a), learned circuits mostly belong to the circuits with short purification times. However, since for longer purification times an exponentially large number of training samples is required, the initial exponential growth is followed by a slow learning curve. Therefore, in Fig.~\ref{Fig:F2}(d), we observe that deep in the pure phase where the majority of circuits have a short purification time, $R_{l}$ asymptotically approaches one.

\section{Complexity results for scrambled initial states and their system-size dependence}
\begin{figure}[]
\centering
\includegraphics[width=7.cm]{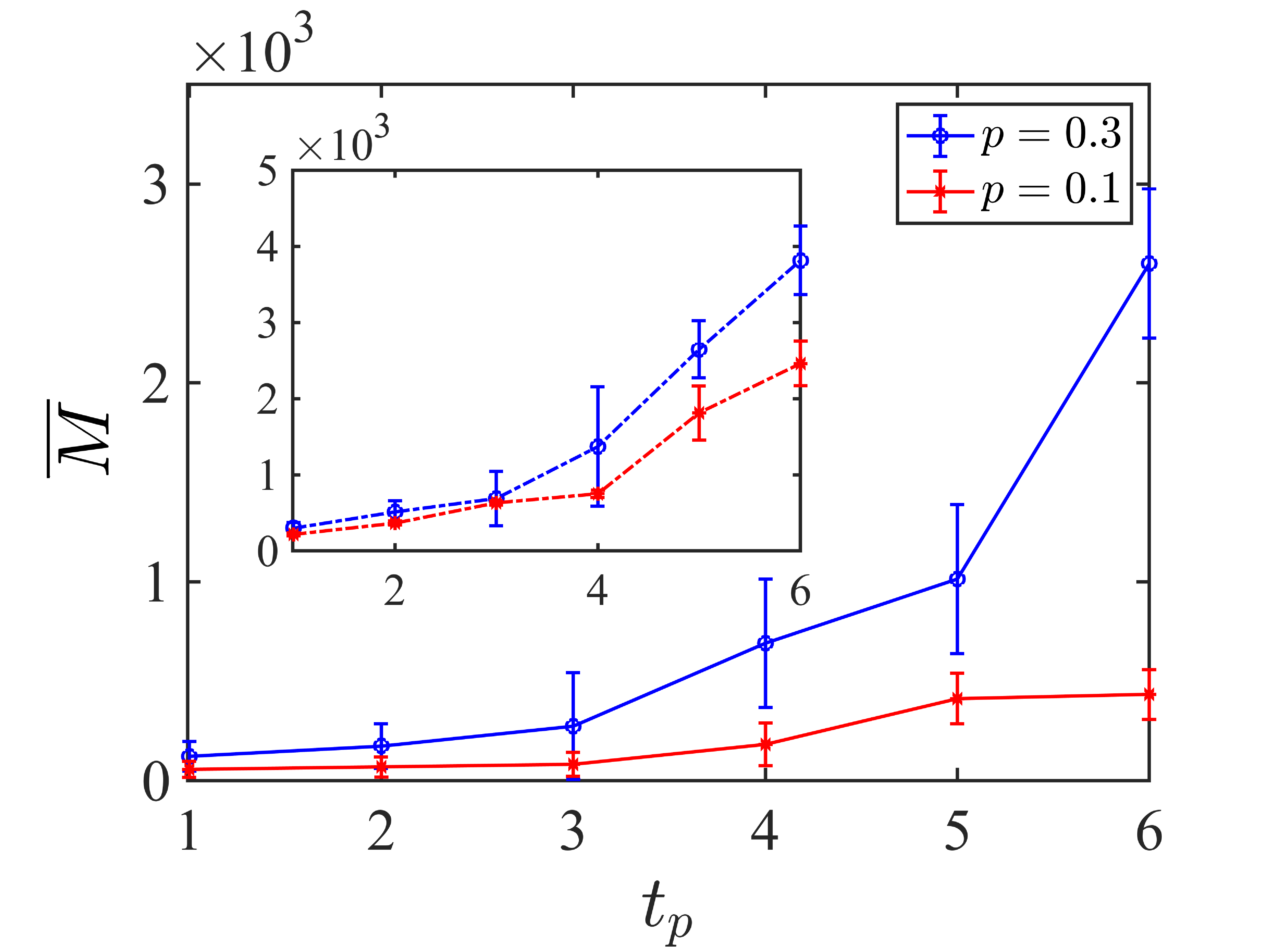}
\caption{Averaged number of training samples required for learning the reference qubit after conditioning on the purification time $t_{p}$, for $p=0.1$ (mixed phase) and $p=0.3$ (pure phase) with a scrambled  initial state. Averaging is performed over $N_c=20$ circuits for each $t_{p}$ and error bars are set according to the standard deviation. We have circuits with $L=128$ qubits. In the main plot measurement outcomes from inside the fixed light-cone box are used for training while for the inset we use the measurement outcomes from the whole circuit. \label{Fig:F5}}
\end{figure}

Here, we present our results for the circuits scrambled by a high-depth random Clifford circuit. Concretely, to obtain such states, we first run our circuits with initial product states only with two-qubit random Clifford gates in the absence of any measurements. This unitary time evolution creates a highly-entangled state after $T\sim L$ time steps with an entanglement entropy proportional to the system size. Next, we entangle the reference qubit to one of the circuit's qubits and run the same circuit in the presence of two-qubit gates and random measurements. As shown in Ref.~\cite{Gullans2020Dynamical}, there is a purification phase transition such that for $p<p_c$ the subsystem entanglement entropy of the circuit after $T\sim L$ still has a volume-law behavior while for $p>p_c$, its entanglement entropy is negligible. Using such initially mixed states, the complexity results are displayed in Fig.~ \ref{Fig:F5}. 
Here, as in Fig.~ \ref{Fig:F4}, we observe a nearly exponential behavior with the purification time. Furthermore, we notice that the conditional learning scheme is more difficult in the pure phase compared to the mixed phase. By comparing the inset and main plots, we also observe that learning with the light-cone data requires less training samples. Finally, by comparing Fig.~\ref{Fig:F2}(b) and Fig.~\ref{Fig:F5} we observe that learning the circuits with scrambled initial conditions requires more training samples than the circuits with product state initial conditions. 

\begin{figure}[]
\centering
\includegraphics[width=8.5cm]{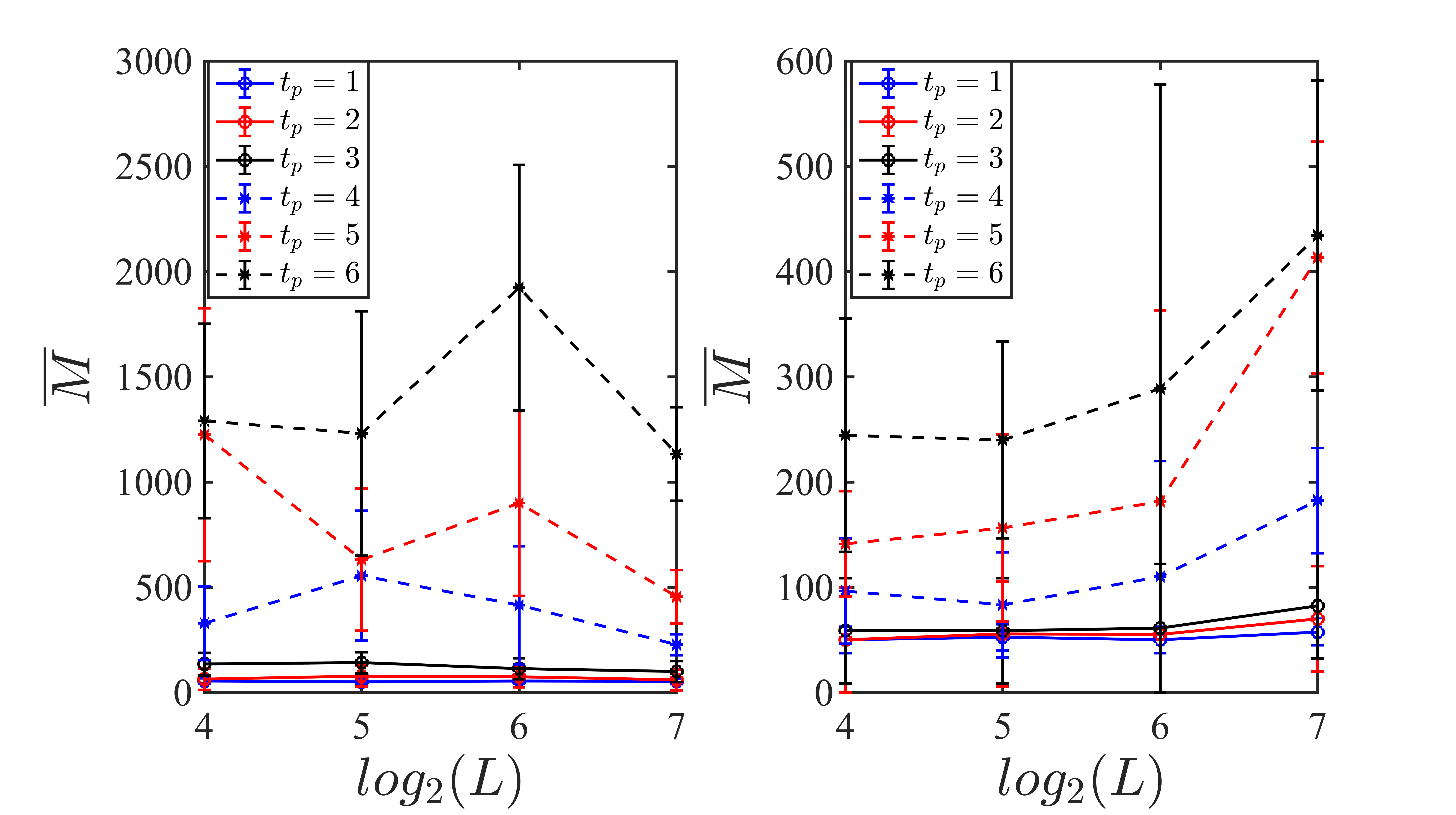}
\caption{Averaged number of training samples required for learning the reference qubit using light-cone data as a function of the system-sizes and different purification times. Averaging is performed over $N_c=20$ circuits for each $t_{p}$ and error bars are set according to the standard deviation. (Left) Results for the area-law phase with $p=0.3$. (Right) Results for the volume-law phase with $p=0.1$. \label{Fig:complexityFuncOfL}}
\end{figure}

Finally, we present further results for the system-size dependence of the sampling complexity of our approach in Fig.\ref{Fig:complexityFuncOfL} where we only use the light-cone measurement outcomes. The $x$ axis represents the system size which includes $L=\{16, 32, 64, 128\}$. Different curves represent different purification times spanning $t_p=\{1, \cdots, 6\}$. In the left panel of this figure we have displayed our results for $p=0.3$ corresponding to the area-law phase and in the right we have displayed our results for the volume-law phase with $p=0.1$. Once the error bars are taken into account, we can see that the sampling complexity is nearly independent of the system size. However, we should note that since the NN decoder that we have employed for these simulations is not necessarily the optimum decoder, we expect some deviation from an exact system-size independent behavior. Changing the system size by a factor of 8, the sample complexity increases by a factor of 2 on average. For more definitive results, we need to consider larger ensembles of circuits with larger $N_c$ and also increase the system size, which would be beyond the scope of this work.


\begin{thebibliography}{70}%
\makeatletter
\providecommand \@ifxundefined [1]{%
 \@ifx{#1\undefined}
}%
\providecommand \@ifnum [1]{%
 \ifnum #1\expandafter \@firstoftwo
 \else \expandafter \@secondoftwo
 \fi
}%
\providecommand \@ifx [1]{%
 \ifx #1\expandafter \@firstoftwo
 \else \expandafter \@secondoftwo
 \fi
}%
\providecommand \natexlab [1]{#1}%
\providecommand \enquote  [1]{``#1''}%
\providecommand \bibnamefont  [1]{#1}%
\providecommand \bibfnamefont [1]{#1}%
\providecommand \citenamefont [1]{#1}%
\providecommand \href@noop [0]{\@secondoftwo}%
\providecommand \href [0]{\begingroup \@sanitize@url \@href}%
\providecommand \@href[1]{\@@startlink{#1}\@@href}%
\providecommand \@@href[1]{\endgroup#1\@@endlink}%
\providecommand \@sanitize@url [0]{\catcode `\\12\catcode `\$12\catcode
  `\&12\catcode `\#12\catcode `\^12\catcode `\_12\catcode `\%12\relax}%
\providecommand \@@startlink[1]{}%
\providecommand \@@endlink[0]{}%
\providecommand \url  [0]{\begingroup\@sanitize@url \@url }%
\providecommand \@url [1]{\endgroup\@href {#1}{\urlprefix }}%
\providecommand \urlprefix  [0]{URL }%
\providecommand \Eprint [0]{\href }%
\providecommand \doibase [0]{http://dx.doi.org/}%
\providecommand \selectlanguage [0]{\@gobble}%
\providecommand \bibinfo  [0]{\@secondoftwo}%
\providecommand \bibfield  [0]{\@secondoftwo}%
\providecommand \translation [1]{[#1]}%
\providecommand \BibitemOpen [0]{}%
\providecommand \bibitemStop [0]{}%
\providecommand \bibitemNoStop [0]{.\EOS\space}%
\providecommand \EOS [0]{\spacefactor3000\relax}%
\providecommand \BibitemShut  [1]{\csname bibitem#1\endcsname}%
\let\auto@bib@innerbib\@empty
%</preamble>
\bibitem [{\citenamefont {Kim}\ and\ \citenamefont
  {Huse}(2013)}]{Huse2013Ballistic}%
  \BibitemOpen
  \bibfield  {author} {\bibinfo {author} {\bibfnamefont {H.}~\bibnamefont
  {Kim}}\ and\ \bibinfo {author} {\bibfnamefont {D.~A.}\ \bibnamefont {Huse}},\
  }\href {\doibase 10.1103/PhysRevLett.111.127205} {\bibfield  {journal}
  {\bibinfo  {journal} {Phys. Rev. Lett.}\ }\textbf {\bibinfo {volume} {111}},\
  \bibinfo {pages} {127205} (\bibinfo {year} {2013})}\BibitemShut {NoStop}%
\bibitem [{\citenamefont {Nandkishore}\ and\ \citenamefont
  {Huse}(2015)}]{Nandkishore2015Many}%
  \BibitemOpen
  \bibfield  {author} {\bibinfo {author} {\bibfnamefont {R.}~\bibnamefont
  {Nandkishore}}\ and\ \bibinfo {author} {\bibfnamefont {D.~A.}\ \bibnamefont
  {Huse}},\ }\href {\doibase 10.1146/annurev-conmatphys-031214-014726}
  {\bibfield  {journal} {\bibinfo  {journal} {Annual Review of Condensed Matter
  Physics}\ }\textbf {\bibinfo {volume} {6}},\ \bibinfo {pages} {15} (\bibinfo
  {year} {2015})}\BibitemShut {NoStop}%
\bibitem [{\citenamefont {Breuer}\ \emph {et~al.}(2002)\citenamefont {Breuer},
  \citenamefont {Petruccione} \emph {et~al.}}]{breuer2002theory}%
  \BibitemOpen
  \bibfield  {author} {\bibinfo {author} {\bibfnamefont {H.-P.}\ \bibnamefont
  {Breuer}}, \bibinfo {author} {\bibfnamefont {F.}~\bibnamefont {Petruccione}},
   \emph {et~al.},\ }\href@noop {} {\emph {\bibinfo {title} {The theory of open
  quantum systems}}}\ (\bibinfo  {publisher} {Oxford University Press on
  Demand},\ \bibinfo {year} {2002})\BibitemShut {NoStop}%
\bibitem [{\citenamefont {Bauer}\ and\ \citenamefont
  {Nayak}(2013)}]{Nayak2013Area}%
  \BibitemOpen
  \bibfield  {author} {\bibinfo {author} {\bibfnamefont {B.}~\bibnamefont
  {Bauer}}\ and\ \bibinfo {author} {\bibfnamefont {C.}~\bibnamefont {Nayak}},\
  }\href {\doibase 10.1088/1742-5468/2013/09/p09005} {\bibfield  {journal}
  {\bibinfo  {journal} {Journal of Statistical Mechanics: Theory and
  Experiment}\ }\textbf {\bibinfo {volume} {2013}},\ \bibinfo {pages} {P09005}
  (\bibinfo {year} {2013})}\BibitemShut {NoStop}%
\bibitem [{\citenamefont {Serbyn}\ \emph {et~al.}(2013)\citenamefont {Serbyn},
  \citenamefont {Papi\ifmmode~\acute{c}\else \'{c}\fi{}},\ and\ \citenamefont
  {Abanin}}]{Abanin2013Local}%
  \BibitemOpen
  \bibfield  {author} {\bibinfo {author} {\bibfnamefont {M.}~\bibnamefont
  {Serbyn}}, \bibinfo {author} {\bibfnamefont {Z.}~\bibnamefont
  {Papi\ifmmode~\acute{c}\else \'{c}\fi{}}}, \ and\ \bibinfo {author}
  {\bibfnamefont {D.~A.}\ \bibnamefont {Abanin}},\ }\href {\doibase
  10.1103/PhysRevLett.111.127201} {\bibfield  {journal} {\bibinfo  {journal}
  {Phys. Rev. Lett.}\ }\textbf {\bibinfo {volume} {111}},\ \bibinfo {pages}
  {127201} (\bibinfo {year} {2013})}\BibitemShut {NoStop}%
\bibitem [{\citenamefont {Skinner}\ \emph {et~al.}(2019)\citenamefont
  {Skinner}, \citenamefont {Ruhman},\ and\ \citenamefont
  {Nahum}}]{Skinner2019Measurement}%
  \BibitemOpen
  \bibfield  {author} {\bibinfo {author} {\bibfnamefont {B.}~\bibnamefont
  {Skinner}}, \bibinfo {author} {\bibfnamefont {J.}~\bibnamefont {Ruhman}}, \
  and\ \bibinfo {author} {\bibfnamefont {A.}~\bibnamefont {Nahum}},\
  }\href@noop {} {\bibfield  {journal} {\bibinfo  {journal} {Phys. Rev. X}\
  }\textbf {\bibinfo {volume} {9}},\ \bibinfo {pages} {031009} (\bibinfo {year}
  {2019})}\BibitemShut {NoStop}%
\bibitem [{\citenamefont {Li}\ \emph {et~al.}(2018)\citenamefont {Li},
  \citenamefont {Chen},\ and\ \citenamefont {Fisher}}]{li2018quantum}%
  \BibitemOpen
  \bibfield  {author} {\bibinfo {author} {\bibfnamefont {Y.}~\bibnamefont
  {Li}}, \bibinfo {author} {\bibfnamefont {X.}~\bibnamefont {Chen}}, \ and\
  \bibinfo {author} {\bibfnamefont {M.~P.~A.}\ \bibnamefont {Fisher}},\ }\href
  {\doibase 10.1103/PhysRevB.98.205136} {\bibfield  {journal} {\bibinfo
  {journal} {Phys. Rev. B}\ }\textbf {\bibinfo {volume} {98}},\ \bibinfo
  {pages} {205136} (\bibinfo {year} {2018})}\BibitemShut {NoStop}%
\bibitem [{\citenamefont {Li}\ \emph {et~al.}(2019)\citenamefont {Li},
  \citenamefont {Chen},\ and\ \citenamefont {Fisher}}]{Fisher2019Measurement}%
  \BibitemOpen
  \bibfield  {author} {\bibinfo {author} {\bibfnamefont {Y.}~\bibnamefont
  {Li}}, \bibinfo {author} {\bibfnamefont {X.}~\bibnamefont {Chen}}, \ and\
  \bibinfo {author} {\bibfnamefont {M.~P.~A.}\ \bibnamefont {Fisher}},\ }\href
  {\doibase 10.1103/PhysRevB.100.134306} {\bibfield  {journal} {\bibinfo
  {journal} {Phys. Rev. B}\ }\textbf {\bibinfo {volume} {100}},\ \bibinfo
  {pages} {134306} (\bibinfo {year} {2019})}\BibitemShut {NoStop}%
\bibitem [{\citenamefont {Noel}\ \emph {et~al.}(2021)\citenamefont {Noel},
  \citenamefont {Niroula}, \citenamefont {Zhu}, \citenamefont {Risinger},
  \citenamefont {Egan}, \citenamefont {Biswas}, \citenamefont {Cetina},
  \citenamefont {Gorshkov}, \citenamefont {Gullans}, \citenamefont {Huse},\
  and\ \citenamefont {Monroe}}]{noel2021observation}%
  \BibitemOpen
  \bibfield  {author} {\bibinfo {author} {\bibfnamefont {C.}~\bibnamefont
  {Noel}}, \bibinfo {author} {\bibfnamefont {P.}~\bibnamefont {Niroula}},
  \bibinfo {author} {\bibfnamefont {D.}~\bibnamefont {Zhu}}, \bibinfo {author}
  {\bibfnamefont {A.}~\bibnamefont {Risinger}}, \bibinfo {author}
  {\bibfnamefont {L.}~\bibnamefont {Egan}}, \bibinfo {author} {\bibfnamefont
  {D.}~\bibnamefont {Biswas}}, \bibinfo {author} {\bibfnamefont
  {M.}~\bibnamefont {Cetina}}, \bibinfo {author} {\bibfnamefont {A.~V.}\
  \bibnamefont {Gorshkov}}, \bibinfo {author} {\bibfnamefont {M.~J.}\
  \bibnamefont {Gullans}}, \bibinfo {author} {\bibfnamefont {D.~A.}\
  \bibnamefont {Huse}}, \ and\ \bibinfo {author} {\bibfnamefont
  {C.}~\bibnamefont {Monroe}},\ }\href@noop {} {\enquote {\bibinfo {title}
  {Measurement-induced quantum phases realized in a trapped-ion quantum
  computer},}\ } (\bibinfo {year} {2021}),\ \Eprint
  {http://arxiv.org/abs/2106.05881} {arXiv:2106.05881 [quant-ph]} \BibitemShut
  {NoStop}%
\bibitem [{\citenamefont {Koh}\ \emph {et~al.}(2022{\natexlab{a}})\citenamefont
  {Koh}, \citenamefont {Sun}, \citenamefont {Motta},\ and\ \citenamefont
  {Minnich}}]{Koh22}%
  \BibitemOpen
  \bibfield  {author} {\bibinfo {author} {\bibfnamefont {J.~M.}\ \bibnamefont
  {Koh}}, \bibinfo {author} {\bibfnamefont {S.-N.}\ \bibnamefont {Sun}},
  \bibinfo {author} {\bibfnamefont {M.}~\bibnamefont {Motta}}, \ and\ \bibinfo
  {author} {\bibfnamefont {A.~J.}\ \bibnamefont {Minnich}},\ }\href@noop {}
  {\bibfield  {journal} {\bibinfo  {journal} {arXiv:2203.04338}\ } (\bibinfo
  {year} {2022}{\natexlab{a}})}\BibitemShut {NoStop}%
\bibitem [{\citenamefont {Gullans}\ and\ \citenamefont
  {Huse}(2020{\natexlab{a}})}]{Gullans2020Dynamical}%
  \BibitemOpen
  \bibfield  {author} {\bibinfo {author} {\bibfnamefont {M.~J.}\ \bibnamefont
  {Gullans}}\ and\ \bibinfo {author} {\bibfnamefont {D.~A.}\ \bibnamefont
  {Huse}},\ }\href {\doibase 10.1103/PhysRevX.10.041020} {\bibfield  {journal}
  {\bibinfo  {journal} {Phys. Rev. X}\ }\textbf {\bibinfo {volume} {10}},\
  \bibinfo {pages} {041020} (\bibinfo {year} {2020}{\natexlab{a}})}\BibitemShut
  {NoStop}%
\bibitem [{\citenamefont {Choi}\ \emph {et~al.}(2020)\citenamefont {Choi},
  \citenamefont {Bao}, \citenamefont {Qi},\ and\ \citenamefont
  {Altman}}]{choi2020quantum}%
  \BibitemOpen
  \bibfield  {author} {\bibinfo {author} {\bibfnamefont {S.}~\bibnamefont
  {Choi}}, \bibinfo {author} {\bibfnamefont {Y.}~\bibnamefont {Bao}}, \bibinfo
  {author} {\bibfnamefont {X.-L.}\ \bibnamefont {Qi}}, \ and\ \bibinfo {author}
  {\bibfnamefont {E.}~\bibnamefont {Altman}},\ }\href@noop {} {\bibfield
  {journal} {\bibinfo  {journal} {Physical Review Letters}\ }\textbf {\bibinfo
  {volume} {125}},\ \bibinfo {pages} {030505} (\bibinfo {year}
  {2020})}\BibitemShut {NoStop}%
\bibitem [{\citenamefont {Jian}\ \emph {et~al.}(2020)\citenamefont {Jian},
  \citenamefont {You}, \citenamefont {Vasseur},\ and\ \citenamefont
  {Ludwig}}]{jian2020measurement}%
  \BibitemOpen
  \bibfield  {author} {\bibinfo {author} {\bibfnamefont {C.-M.}\ \bibnamefont
  {Jian}}, \bibinfo {author} {\bibfnamefont {Y.-Z.}\ \bibnamefont {You}},
  \bibinfo {author} {\bibfnamefont {R.}~\bibnamefont {Vasseur}}, \ and\
  \bibinfo {author} {\bibfnamefont {A.~W.}\ \bibnamefont {Ludwig}},\
  }\href@noop {} {\bibfield  {journal} {\bibinfo  {journal} {Physical Review
  B}\ }\textbf {\bibinfo {volume} {101}},\ \bibinfo {pages} {104302} (\bibinfo
  {year} {2020})}\BibitemShut {NoStop}%
\bibitem [{\citenamefont {Bao}\ \emph {et~al.}(2020)\citenamefont {Bao},
  \citenamefont {Choi},\ and\ \citenamefont {Altman}}]{bao2020theory}%
  \BibitemOpen
  \bibfield  {author} {\bibinfo {author} {\bibfnamefont {Y.}~\bibnamefont
  {Bao}}, \bibinfo {author} {\bibfnamefont {S.}~\bibnamefont {Choi}}, \ and\
  \bibinfo {author} {\bibfnamefont {E.}~\bibnamefont {Altman}},\ }\href@noop {}
  {\bibfield  {journal} {\bibinfo  {journal} {Physical Review B}\ }\textbf
  {\bibinfo {volume} {101}},\ \bibinfo {pages} {104301} (\bibinfo {year}
  {2020})}\BibitemShut {NoStop}%
\bibitem [{\citenamefont {Zabalo}\ \emph {et~al.}(2020)\citenamefont {Zabalo},
  \citenamefont {Gullans}, \citenamefont {Wilson}, \citenamefont
  {Gopalakrishnan}, \citenamefont {Huse},\ and\ \citenamefont
  {Pixley}}]{zabalo2020critical}%
  \BibitemOpen
  \bibfield  {author} {\bibinfo {author} {\bibfnamefont {A.}~\bibnamefont
  {Zabalo}}, \bibinfo {author} {\bibfnamefont {M.~J.}\ \bibnamefont {Gullans}},
  \bibinfo {author} {\bibfnamefont {J.~H.}\ \bibnamefont {Wilson}}, \bibinfo
  {author} {\bibfnamefont {S.}~\bibnamefont {Gopalakrishnan}}, \bibinfo
  {author} {\bibfnamefont {D.~A.}\ \bibnamefont {Huse}}, \ and\ \bibinfo
  {author} {\bibfnamefont {J.}~\bibnamefont {Pixley}},\ }\href@noop {}
  {\bibfield  {journal} {\bibinfo  {journal} {Physical Review B}\ }\textbf
  {\bibinfo {volume} {101}},\ \bibinfo {pages} {060301} (\bibinfo {year}
  {2020})}\BibitemShut {NoStop}%
\bibitem [{\citenamefont {Tang}\ and\ \citenamefont
  {Zhu}(2020)}]{Tang2020Measurement}%
  \BibitemOpen
  \bibfield  {author} {\bibinfo {author} {\bibfnamefont {Q.}~\bibnamefont
  {Tang}}\ and\ \bibinfo {author} {\bibfnamefont {W.}~\bibnamefont {Zhu}},\
  }\href {\doibase 10.1103/PhysRevResearch.2.013022} {\bibfield  {journal}
  {\bibinfo  {journal} {Phys. Rev. Research}\ }\textbf {\bibinfo {volume}
  {2}},\ \bibinfo {pages} {013022} (\bibinfo {year} {2020})}\BibitemShut
  {NoStop}%
\bibitem [{\citenamefont {Fuji}\ and\ \citenamefont
  {Ashida}(2020)}]{Fuji2020Measurement}%
  \BibitemOpen
  \bibfield  {author} {\bibinfo {author} {\bibfnamefont {Y.}~\bibnamefont
  {Fuji}}\ and\ \bibinfo {author} {\bibfnamefont {Y.}~\bibnamefont {Ashida}},\
  }\href {\doibase 10.1103/PhysRevB.102.054302} {\bibfield  {journal} {\bibinfo
   {journal} {Phys. Rev. B}\ }\textbf {\bibinfo {volume} {102}},\ \bibinfo
  {pages} {054302} (\bibinfo {year} {2020})}\BibitemShut {NoStop}%
\bibitem [{\citenamefont {Turkeshi}\ \emph {et~al.}(2020)\citenamefont
  {Turkeshi}, \citenamefont {Fazio},\ and\ \citenamefont
  {Dalmonte}}]{Turkeshi2020Measurement}%
  \BibitemOpen
  \bibfield  {author} {\bibinfo {author} {\bibfnamefont {X.}~\bibnamefont
  {Turkeshi}}, \bibinfo {author} {\bibfnamefont {R.}~\bibnamefont {Fazio}}, \
  and\ \bibinfo {author} {\bibfnamefont {M.}~\bibnamefont {Dalmonte}},\ }\href
  {\doibase 10.1103/PhysRevB.102.014315} {\bibfield  {journal} {\bibinfo
  {journal} {Phys. Rev. B}\ }\textbf {\bibinfo {volume} {102}},\ \bibinfo
  {pages} {014315} (\bibinfo {year} {2020})}\BibitemShut {NoStop}%
\bibitem [{\citenamefont {Ippoliti}\ \emph {et~al.}(2021)\citenamefont
  {Ippoliti}, \citenamefont {Gullans}, \citenamefont {Gopalakrishnan},
  \citenamefont {Huse},\ and\ \citenamefont
  {Khemani}}]{Ippoliti2021Entanglement}%
  \BibitemOpen
  \bibfield  {author} {\bibinfo {author} {\bibfnamefont {M.}~\bibnamefont
  {Ippoliti}}, \bibinfo {author} {\bibfnamefont {M.~J.}\ \bibnamefont
  {Gullans}}, \bibinfo {author} {\bibfnamefont {S.}~\bibnamefont
  {Gopalakrishnan}}, \bibinfo {author} {\bibfnamefont {D.~A.}\ \bibnamefont
  {Huse}}, \ and\ \bibinfo {author} {\bibfnamefont {V.}~\bibnamefont
  {Khemani}},\ }\href {\doibase 10.1103/PhysRevX.11.011030} {\bibfield
  {journal} {\bibinfo  {journal} {Phys. Rev. X}\ }\textbf {\bibinfo {volume}
  {11}},\ \bibinfo {pages} {011030} (\bibinfo {year} {2021})}\BibitemShut
  {NoStop}%
\bibitem [{\citenamefont {Lavasani}\ \emph {et~al.}(2021)\citenamefont
  {Lavasani}, \citenamefont {Alavirad},\ and\ \citenamefont
  {Barkeshli}}]{Lavasani2021Measurement}%
  \BibitemOpen
  \bibfield  {author} {\bibinfo {author} {\bibfnamefont {A.}~\bibnamefont
  {Lavasani}}, \bibinfo {author} {\bibfnamefont {Y.}~\bibnamefont {Alavirad}},
  \ and\ \bibinfo {author} {\bibfnamefont {M.}~\bibnamefont {Barkeshli}},\
  }\href {\doibase 10.1038/s41567-020-01112-z} {\bibfield  {journal} {\bibinfo
  {journal} {Nature Physics}\ }\textbf {\bibinfo {volume} {17}},\ \bibinfo
  {pages} {342} (\bibinfo {year} {2021})}\BibitemShut {NoStop}%
\bibitem [{\citenamefont {Sang}\ and\ \citenamefont
  {Hsieh}(2021)}]{Hsieh2021Measurement}%
  \BibitemOpen
  \bibfield  {author} {\bibinfo {author} {\bibfnamefont {S.}~\bibnamefont
  {Sang}}\ and\ \bibinfo {author} {\bibfnamefont {T.~H.}\ \bibnamefont
  {Hsieh}},\ }\href {\doibase 10.1103/PhysRevResearch.3.023200} {\bibfield
  {journal} {\bibinfo  {journal} {Phys. Rev. Research}\ }\textbf {\bibinfo
  {volume} {3}},\ \bibinfo {pages} {023200} (\bibinfo {year}
  {2021})}\BibitemShut {NoStop}%
\bibitem [{\citenamefont {Van~Regemortel}\ \emph {et~al.}(2021)\citenamefont
  {Van~Regemortel}, \citenamefont {Cian}, \citenamefont {Seif}, \citenamefont
  {Dehghani},\ and\ \citenamefont {Hafezi}}]{Mathias2021Entanglement}%
  \BibitemOpen
  \bibfield  {author} {\bibinfo {author} {\bibfnamefont {M.}~\bibnamefont
  {Van~Regemortel}}, \bibinfo {author} {\bibfnamefont {Z.-P.}\ \bibnamefont
  {Cian}}, \bibinfo {author} {\bibfnamefont {A.}~\bibnamefont {Seif}}, \bibinfo
  {author} {\bibfnamefont {H.}~\bibnamefont {Dehghani}}, \ and\ \bibinfo
  {author} {\bibfnamefont {M.}~\bibnamefont {Hafezi}},\ }\href {\doibase
  10.1103/PhysRevLett.126.123604} {\bibfield  {journal} {\bibinfo  {journal}
  {Phys. Rev. Lett.}\ }\textbf {\bibinfo {volume} {126}},\ \bibinfo {pages}
  {123604} (\bibinfo {year} {2021})}\BibitemShut {NoStop}%
\bibitem [{\citenamefont {Buchhold}\ \emph {et~al.}(2021)\citenamefont
  {Buchhold}, \citenamefont {Minoguchi}, \citenamefont {Altland},\ and\
  \citenamefont {Diehl}}]{Buchhold2021Effective}%
  \BibitemOpen
  \bibfield  {author} {\bibinfo {author} {\bibfnamefont {M.}~\bibnamefont
  {Buchhold}}, \bibinfo {author} {\bibfnamefont {Y.}~\bibnamefont {Minoguchi}},
  \bibinfo {author} {\bibfnamefont {A.}~\bibnamefont {Altland}}, \ and\
  \bibinfo {author} {\bibfnamefont {S.}~\bibnamefont {Diehl}},\ }\href
  {\doibase 10.1103/PhysRevX.11.041004} {\bibfield  {journal} {\bibinfo
  {journal} {Phys. Rev. X}\ }\textbf {\bibinfo {volume} {11}},\ \bibinfo
  {pages} {041004} (\bibinfo {year} {2021})}\BibitemShut {NoStop}%
\bibitem [{\citenamefont {Bao}\ \emph {et~al.}(2021)\citenamefont {Bao},
  \citenamefont {Choi},\ and\ \citenamefont {Altman}}]{Altman2021Symmetry}%
  \BibitemOpen
  \bibfield  {author} {\bibinfo {author} {\bibfnamefont {Y.}~\bibnamefont
  {Bao}}, \bibinfo {author} {\bibfnamefont {S.}~\bibnamefont {Choi}}, \ and\
  \bibinfo {author} {\bibfnamefont {E.}~\bibnamefont {Altman}},\ }\href
  {\doibase https://doi.org/10.1016/j.aop.2021.168618} {\bibfield  {journal}
  {\bibinfo  {journal} {Annals of Physics}\ }\textbf {\bibinfo {volume}
  {435}},\ \bibinfo {pages} {168618} (\bibinfo {year} {2021})},\ \bibinfo
  {note} {special issue on Philip W. Anderson}\BibitemShut {NoStop}%
\bibitem [{\citenamefont {Jian}\ \emph {et~al.}(2021)\citenamefont {Jian},
  \citenamefont {Liu}, \citenamefont {Chen}, \citenamefont {Swingle},\ and\
  \citenamefont {Zhang}}]{Jian2021Measurement}%
  \BibitemOpen
  \bibfield  {author} {\bibinfo {author} {\bibfnamefont {S.-K.}\ \bibnamefont
  {Jian}}, \bibinfo {author} {\bibfnamefont {C.}~\bibnamefont {Liu}}, \bibinfo
  {author} {\bibfnamefont {X.}~\bibnamefont {Chen}}, \bibinfo {author}
  {\bibfnamefont {B.}~\bibnamefont {Swingle}}, \ and\ \bibinfo {author}
  {\bibfnamefont {P.}~\bibnamefont {Zhang}},\ }\href {\doibase
  10.1103/PhysRevLett.127.140601} {\bibfield  {journal} {\bibinfo  {journal}
  {Phys. Rev. Lett.}\ }\textbf {\bibinfo {volume} {127}},\ \bibinfo {pages}
  {140601} (\bibinfo {year} {2021})}\BibitemShut {NoStop}%
\bibitem [{\citenamefont {Czischek}\ \emph {et~al.}(2021)\citenamefont
  {Czischek}, \citenamefont {Torlai}, \citenamefont {Ray}, \citenamefont
  {Islam},\ and\ \citenamefont {Melko}}]{Czischek21}%
  \BibitemOpen
  \bibfield  {author} {\bibinfo {author} {\bibfnamefont {S.}~\bibnamefont
  {Czischek}}, \bibinfo {author} {\bibfnamefont {G.}~\bibnamefont {Torlai}},
  \bibinfo {author} {\bibfnamefont {S.}~\bibnamefont {Ray}}, \bibinfo {author}
  {\bibfnamefont {R.}~\bibnamefont {Islam}}, \ and\ \bibinfo {author}
  {\bibfnamefont {R.~G.}\ \bibnamefont {Melko}},\ }\href {\doibase
  10.1103/PhysRevA.104.062405} {\bibfield  {journal} {\bibinfo  {journal}
  {Phys. Rev. A}\ }\textbf {\bibinfo {volume} {104}},\ \bibinfo {pages}
  {062405} (\bibinfo {year} {2021})}\BibitemShut {NoStop}%
\bibitem [{\citenamefont {Potter}\ and\ \citenamefont
  {Vasseur}(2021)}]{potter2021entanglement}%
  \BibitemOpen
  \bibfield  {author} {\bibinfo {author} {\bibfnamefont {A.~C.}\ \bibnamefont
  {Potter}}\ and\ \bibinfo {author} {\bibfnamefont {R.}~\bibnamefont
  {Vasseur}},\ }\href@noop {} {\bibfield  {journal} {\bibinfo  {journal} {arXiv
  preprint arXiv:2111.08018}\ } (\bibinfo {year} {2021})}\BibitemShut {NoStop}%
\bibitem [{\citenamefont {Turkeshi}(2021)}]{turkeshi2021measurement}%
  \BibitemOpen
  \bibfield  {author} {\bibinfo {author} {\bibfnamefont {X.}~\bibnamefont
  {Turkeshi}},\ }\href@noop {} {\bibfield  {journal} {\bibinfo  {journal}
  {arXiv preprint arXiv:2101.06245}\ } (\bibinfo {year} {2021})}\BibitemShut
  {NoStop}%
\bibitem [{\citenamefont {Block}\ \emph {et~al.}(2022)\citenamefont {Block},
  \citenamefont {Bao}, \citenamefont {Choi}, \citenamefont {Altman},\ and\
  \citenamefont {Yao}}]{Altman2022Measurement}%
  \BibitemOpen
  \bibfield  {author} {\bibinfo {author} {\bibfnamefont {M.}~\bibnamefont
  {Block}}, \bibinfo {author} {\bibfnamefont {Y.}~\bibnamefont {Bao}}, \bibinfo
  {author} {\bibfnamefont {S.}~\bibnamefont {Choi}}, \bibinfo {author}
  {\bibfnamefont {E.}~\bibnamefont {Altman}}, \ and\ \bibinfo {author}
  {\bibfnamefont {N.~Y.}\ \bibnamefont {Yao}},\ }\href {\doibase
  10.1103/PhysRevLett.128.010604} {\bibfield  {journal} {\bibinfo  {journal}
  {Phys. Rev. Lett.}\ }\textbf {\bibinfo {volume} {128}},\ \bibinfo {pages}
  {010604} (\bibinfo {year} {2022})}\BibitemShut {NoStop}%
\bibitem [{\citenamefont {Minato}\ \emph {et~al.}(2022)\citenamefont {Minato},
  \citenamefont {Sugimoto}, \citenamefont {Kuwahara},\ and\ \citenamefont
  {Saito}}]{Minato2022Fate}%
  \BibitemOpen
  \bibfield  {author} {\bibinfo {author} {\bibfnamefont {T.}~\bibnamefont
  {Minato}}, \bibinfo {author} {\bibfnamefont {K.}~\bibnamefont {Sugimoto}},
  \bibinfo {author} {\bibfnamefont {T.}~\bibnamefont {Kuwahara}}, \ and\
  \bibinfo {author} {\bibfnamefont {K.}~\bibnamefont {Saito}},\ }\href
  {\doibase 10.1103/PhysRevLett.128.010603} {\bibfield  {journal} {\bibinfo
  {journal} {Phys. Rev. Lett.}\ }\textbf {\bibinfo {volume} {128}},\ \bibinfo
  {pages} {010603} (\bibinfo {year} {2022})}\BibitemShut {NoStop}%
\bibitem [{\citenamefont {M\"uller}\ \emph {et~al.}(2022)\citenamefont
  {M\"uller}, \citenamefont {Diehl},\ and\ \citenamefont
  {Buchhold}}]{Muller2022Measurement}%
  \BibitemOpen
  \bibfield  {author} {\bibinfo {author} {\bibfnamefont {T.}~\bibnamefont
  {M\"uller}}, \bibinfo {author} {\bibfnamefont {S.}~\bibnamefont {Diehl}}, \
  and\ \bibinfo {author} {\bibfnamefont {M.}~\bibnamefont {Buchhold}},\ }\href
  {\doibase 10.1103/PhysRevLett.128.010605} {\bibfield  {journal} {\bibinfo
  {journal} {Phys. Rev. Lett.}\ }\textbf {\bibinfo {volume} {128}},\ \bibinfo
  {pages} {010605} (\bibinfo {year} {2022})}\BibitemShut {NoStop}%
\bibitem [{\citenamefont {Van~Regemortel}\ \emph {et~al.}(2022)\citenamefont
  {Van~Regemortel}, \citenamefont {Shtanko}, \citenamefont {Garcia-Pintos},
  \citenamefont {Deshpande}, \citenamefont {Dehghani}, \citenamefont
  {Gorshkov},\ and\ \citenamefont {Hafezi}}]{van2022monitoring}%
  \BibitemOpen
  \bibfield  {author} {\bibinfo {author} {\bibfnamefont {M.}~\bibnamefont
  {Van~Regemortel}}, \bibinfo {author} {\bibfnamefont {O.}~\bibnamefont
  {Shtanko}}, \bibinfo {author} {\bibfnamefont {L.~P.}\ \bibnamefont
  {Garcia-Pintos}}, \bibinfo {author} {\bibfnamefont {A.}~\bibnamefont
  {Deshpande}}, \bibinfo {author} {\bibfnamefont {H.}~\bibnamefont {Dehghani}},
  \bibinfo {author} {\bibfnamefont {A.~V.}\ \bibnamefont {Gorshkov}}, \ and\
  \bibinfo {author} {\bibfnamefont {M.}~\bibnamefont {Hafezi}},\ }\href@noop {}
  {\bibfield  {journal} {\bibinfo  {journal} {arXiv preprint arXiv:2201.12672}\
  } (\bibinfo {year} {2022})}\BibitemShut {NoStop}%
\bibitem [{\citenamefont {Koh}\ \emph {et~al.}(2022{\natexlab{b}})\citenamefont
  {Koh}, \citenamefont {Sun}, \citenamefont {Motta},\ and\ \citenamefont
  {Minnich}}]{koh2022experimental}%
  \BibitemOpen
  \bibfield  {author} {\bibinfo {author} {\bibfnamefont {J.~M.}\ \bibnamefont
  {Koh}}, \bibinfo {author} {\bibfnamefont {S.-N.}\ \bibnamefont {Sun}},
  \bibinfo {author} {\bibfnamefont {M.}~\bibnamefont {Motta}}, \ and\ \bibinfo
  {author} {\bibfnamefont {A.~J.}\ \bibnamefont {Minnich}},\ }\href@noop {}
  {\bibfield  {journal} {\bibinfo  {journal} {arXiv preprint arXiv:2203.04338}\
  } (\bibinfo {year} {2022}{\natexlab{b}})}\BibitemShut {NoStop}%
\bibitem [{\citenamefont {Gullans}\ \emph {et~al.}(2021)\citenamefont
  {Gullans}, \citenamefont {Krastanov}, \citenamefont {Huse}, \citenamefont
  {Jiang},\ and\ \citenamefont {Flammia}}]{Gullans2021Quantum}%
  \BibitemOpen
  \bibfield  {author} {\bibinfo {author} {\bibfnamefont {M.~J.}\ \bibnamefont
  {Gullans}}, \bibinfo {author} {\bibfnamefont {S.}~\bibnamefont {Krastanov}},
  \bibinfo {author} {\bibfnamefont {D.~A.}\ \bibnamefont {Huse}}, \bibinfo
  {author} {\bibfnamefont {L.}~\bibnamefont {Jiang}}, \ and\ \bibinfo {author}
  {\bibfnamefont {S.~T.}\ \bibnamefont {Flammia}},\ }\href {\doibase
  10.1103/PhysRevX.11.031066} {\bibfield  {journal} {\bibinfo  {journal} {Phys.
  Rev. X}\ }\textbf {\bibinfo {volume} {11}},\ \bibinfo {pages} {031066}
  (\bibinfo {year} {2021})}\BibitemShut {NoStop}%
\bibitem [{\citenamefont {Fan}\ \emph {et~al.}(2021)\citenamefont {Fan},
  \citenamefont {Vijay}, \citenamefont {Vishwanath},\ and\ \citenamefont
  {You}}]{fan2021self}%
  \BibitemOpen
  \bibfield  {author} {\bibinfo {author} {\bibfnamefont {R.}~\bibnamefont
  {Fan}}, \bibinfo {author} {\bibfnamefont {S.}~\bibnamefont {Vijay}}, \bibinfo
  {author} {\bibfnamefont {A.}~\bibnamefont {Vishwanath}}, \ and\ \bibinfo
  {author} {\bibfnamefont {Y.-Z.}\ \bibnamefont {You}},\ }\href@noop {}
  {\bibfield  {journal} {\bibinfo  {journal} {Physical Review B}\ }\textbf
  {\bibinfo {volume} {103}},\ \bibinfo {pages} {174309} (\bibinfo {year}
  {2021})}\BibitemShut {NoStop}%
\bibitem [{\citenamefont {Li}\ and\ \citenamefont
  {Fisher}(2021)}]{Li2021Statistical}%
  \BibitemOpen
  \bibfield  {author} {\bibinfo {author} {\bibfnamefont {Y.}~\bibnamefont
  {Li}}\ and\ \bibinfo {author} {\bibfnamefont {M.~P.~A.}\ \bibnamefont
  {Fisher}},\ }\href {\doibase 10.1103/PhysRevB.103.104306} {\bibfield
  {journal} {\bibinfo  {journal} {Phys. Rev. B}\ }\textbf {\bibinfo {volume}
  {103}},\ \bibinfo {pages} {104306} (\bibinfo {year} {2021})}\BibitemShut
  {NoStop}%
\bibitem [{\citenamefont {Yoshida}(2021)}]{yoshida2021decoding}%
  \BibitemOpen
  \bibfield  {author} {\bibinfo {author} {\bibfnamefont {B.}~\bibnamefont
  {Yoshida}},\ }\href@noop {} {\bibfield  {journal} {\bibinfo  {journal} {arXiv
  preprint arXiv:2109.08691}\ } (\bibinfo {year} {2021})}\BibitemShut {NoStop}%
\bibitem [{\citenamefont {Gullans}\ and\ \citenamefont
  {Huse}(2020{\natexlab{b}})}]{Gullans2020Scalable}%
  \BibitemOpen
  \bibfield  {author} {\bibinfo {author} {\bibfnamefont {M.~J.}\ \bibnamefont
  {Gullans}}\ and\ \bibinfo {author} {\bibfnamefont {D.~A.}\ \bibnamefont
  {Huse}},\ }\href {\doibase 10.1103/PhysRevLett.125.070606} {\bibfield
  {journal} {\bibinfo  {journal} {Phys. Rev. Lett.}\ }\textbf {\bibinfo
  {volume} {125}},\ \bibinfo {pages} {070606} (\bibinfo {year}
  {2020}{\natexlab{b}})}\BibitemShut {NoStop}%
\bibitem [{\citenamefont {Carrasquilla}(2020)}]{carrasquilla2020machine}%
  \BibitemOpen
  \bibfield  {author} {\bibinfo {author} {\bibfnamefont {J.}~\bibnamefont
  {Carrasquilla}},\ }\href@noop {} {\bibfield  {journal} {\bibinfo  {journal}
  {Advances in Physics: X}\ }\textbf {\bibinfo {volume} {5}},\ \bibinfo {pages}
  {1797528} (\bibinfo {year} {2020})}\BibitemShut {NoStop}%
\bibitem [{\citenamefont {Torlai}\ and\ \citenamefont
  {Melko}(2017)}]{Torlai2017Neural}%
  \BibitemOpen
  \bibfield  {author} {\bibinfo {author} {\bibfnamefont {G.}~\bibnamefont
  {Torlai}}\ and\ \bibinfo {author} {\bibfnamefont {R.~G.}\ \bibnamefont
  {Melko}},\ }\href {\doibase 10.1103/PhysRevLett.119.030501} {\bibfield
  {journal} {\bibinfo  {journal} {Phys. Rev. Lett.}\ }\textbf {\bibinfo
  {volume} {119}},\ \bibinfo {pages} {030501} (\bibinfo {year}
  {2017})}\BibitemShut {NoStop}%
\bibitem [{\citenamefont {Krastanov}\ and\ \citenamefont
  {Jiang}(2017)}]{krastanov2017deep}%
  \BibitemOpen
  \bibfield  {author} {\bibinfo {author} {\bibfnamefont {S.}~\bibnamefont
  {Krastanov}}\ and\ \bibinfo {author} {\bibfnamefont {L.}~\bibnamefont
  {Jiang}},\ }\href@noop {} {\bibfield  {journal} {\bibinfo  {journal}
  {Scientific reports}\ }\textbf {\bibinfo {volume} {7}},\ \bibinfo {pages} {1}
  (\bibinfo {year} {2017})}\BibitemShut {NoStop}%
\bibitem [{\citenamefont {Baireuther}\ \emph {et~al.}(2018)\citenamefont
  {Baireuther}, \citenamefont {O'Brien}, \citenamefont {Tarasinski},\ and\
  \citenamefont {Beenakker}}]{baireuther2018machine}%
  \BibitemOpen
  \bibfield  {author} {\bibinfo {author} {\bibfnamefont {P.}~\bibnamefont
  {Baireuther}}, \bibinfo {author} {\bibfnamefont {T.~E.}\ \bibnamefont
  {O'Brien}}, \bibinfo {author} {\bibfnamefont {B.}~\bibnamefont {Tarasinski}},
  \ and\ \bibinfo {author} {\bibfnamefont {C.~W.}\ \bibnamefont {Beenakker}},\
  }\href@noop {} {\bibfield  {journal} {\bibinfo  {journal} {Quantum}\ }\textbf
  {\bibinfo {volume} {2}},\ \bibinfo {pages} {48} (\bibinfo {year}
  {2018})}\BibitemShut {NoStop}%
\bibitem [{\citenamefont {Andreasson}\ \emph {et~al.}(2019)\citenamefont
  {Andreasson}, \citenamefont {Johansson}, \citenamefont {Liljestrand},\ and\
  \citenamefont {Granath}}]{andreasson2019quantum}%
  \BibitemOpen
  \bibfield  {author} {\bibinfo {author} {\bibfnamefont {P.}~\bibnamefont
  {Andreasson}}, \bibinfo {author} {\bibfnamefont {J.}~\bibnamefont
  {Johansson}}, \bibinfo {author} {\bibfnamefont {S.}~\bibnamefont
  {Liljestrand}}, \ and\ \bibinfo {author} {\bibfnamefont {M.}~\bibnamefont
  {Granath}},\ }\href@noop {} {\bibfield  {journal} {\bibinfo  {journal}
  {Quantum}\ }\textbf {\bibinfo {volume} {3}},\ \bibinfo {pages} {183}
  (\bibinfo {year} {2019})}\BibitemShut {NoStop}%
\bibitem [{\citenamefont {Nautrup}\ \emph {et~al.}(2019)\citenamefont
  {Nautrup}, \citenamefont {Delfosse}, \citenamefont {Dunjko}, \citenamefont
  {Briegel},\ and\ \citenamefont {Friis}}]{nautrup2019optimizing}%
  \BibitemOpen
  \bibfield  {author} {\bibinfo {author} {\bibfnamefont {H.~P.}\ \bibnamefont
  {Nautrup}}, \bibinfo {author} {\bibfnamefont {N.}~\bibnamefont {Delfosse}},
  \bibinfo {author} {\bibfnamefont {V.}~\bibnamefont {Dunjko}}, \bibinfo
  {author} {\bibfnamefont {H.~J.}\ \bibnamefont {Briegel}}, \ and\ \bibinfo
  {author} {\bibfnamefont {N.}~\bibnamefont {Friis}},\ }\href@noop {}
  {\bibfield  {journal} {\bibinfo  {journal} {Quantum}\ }\textbf {\bibinfo
  {volume} {3}},\ \bibinfo {pages} {215} (\bibinfo {year} {2019})}\BibitemShut
  {NoStop}%
\bibitem [{\citenamefont {Liu}\ and\ \citenamefont
  {Poulin}(2019)}]{Liu2019Neural}%
  \BibitemOpen
  \bibfield  {author} {\bibinfo {author} {\bibfnamefont {Y.-H.}\ \bibnamefont
  {Liu}}\ and\ \bibinfo {author} {\bibfnamefont {D.}~\bibnamefont {Poulin}},\
  }\href {\doibase 10.1103/PhysRevLett.122.200501} {\bibfield  {journal}
  {\bibinfo  {journal} {Phys. Rev. Lett.}\ }\textbf {\bibinfo {volume} {122}},\
  \bibinfo {pages} {200501} (\bibinfo {year} {2019})}\BibitemShut {NoStop}%
\bibitem [{\citenamefont {Flurin}\ \emph {et~al.}(2020)\citenamefont {Flurin},
  \citenamefont {Martin}, \citenamefont {Hacohen-Gourgy},\ and\ \citenamefont
  {Siddiqi}}]{Siddiqi2020Using}%
  \BibitemOpen
  \bibfield  {author} {\bibinfo {author} {\bibfnamefont {E.}~\bibnamefont
  {Flurin}}, \bibinfo {author} {\bibfnamefont {L.~S.}\ \bibnamefont {Martin}},
  \bibinfo {author} {\bibfnamefont {S.}~\bibnamefont {Hacohen-Gourgy}}, \ and\
  \bibinfo {author} {\bibfnamefont {I.}~\bibnamefont {Siddiqi}},\ }\href
  {\doibase 10.1103/PhysRevX.10.011006} {\bibfield  {journal} {\bibinfo
  {journal} {Phys. Rev. X}\ }\textbf {\bibinfo {volume} {10}},\ \bibinfo
  {pages} {011006} (\bibinfo {year} {2020})}\BibitemShut {NoStop}%
\bibitem [{\citenamefont {Sweke}\ \emph {et~al.}(2020)\citenamefont {Sweke},
  \citenamefont {Kesselring}, \citenamefont {van Nieuwenburg},\ and\
  \citenamefont {Eisert}}]{sweke2020reinforcement}%
  \BibitemOpen
  \bibfield  {author} {\bibinfo {author} {\bibfnamefont {R.}~\bibnamefont
  {Sweke}}, \bibinfo {author} {\bibfnamefont {M.~S.}\ \bibnamefont
  {Kesselring}}, \bibinfo {author} {\bibfnamefont {E.~P.}\ \bibnamefont {van
  Nieuwenburg}}, \ and\ \bibinfo {author} {\bibfnamefont {J.}~\bibnamefont
  {Eisert}},\ }\href@noop {} {\bibfield  {journal} {\bibinfo  {journal}
  {Machine Learning: Science and Technology}\ }\textbf {\bibinfo {volume}
  {2}},\ \bibinfo {pages} {025005} (\bibinfo {year} {2020})}\BibitemShut
  {NoStop}%
\bibitem [{\citenamefont {Schumacher}\ and\ \citenamefont
  {Nielsen}(1996)}]{Schumacher96}%
  \BibitemOpen
  \bibfield  {author} {\bibinfo {author} {\bibfnamefont {B.}~\bibnamefont
  {Schumacher}}\ and\ \bibinfo {author} {\bibfnamefont {M.~A.}\ \bibnamefont
  {Nielsen}},\ }\href {\doibase 10.1103/PhysRevA.54.2629} {\bibfield  {journal}
  {\bibinfo  {journal} {Phys. Rev. A}\ }\textbf {\bibinfo {volume} {54}},\
  \bibinfo {pages} {2629} (\bibinfo {year} {1996})}\BibitemShut {NoStop}%
\bibitem [{\citenamefont {Niculescu-Mizil}\ and\ \citenamefont
  {Caruana}(2005)}]{Niculescu2005Predicting}%
  \BibitemOpen
  \bibfield  {author} {\bibinfo {author} {\bibfnamefont {A.}~\bibnamefont
  {Niculescu-Mizil}}\ and\ \bibinfo {author} {\bibfnamefont {R.}~\bibnamefont
  {Caruana}},\ }in\ \href {\doibase 10.1145/1102351.1102430} {\emph {\bibinfo
  {booktitle} {Proceedings of the 22nd International Conference on Machine
  Learning}}},\ \bibinfo {series and number} {ICML '05}\ (\bibinfo  {publisher}
  {Association for Computing Machinery},\ \bibinfo {address} {New York, NY,
  USA},\ \bibinfo {year} {2005})\ p.\ \bibinfo {pages} {625–632}\BibitemShut
  {NoStop}%
\bibitem [{\citenamefont {Guo}\ \emph {et~al.}(2017)\citenamefont {Guo},
  \citenamefont {Pleiss}, \citenamefont {Sun},\ and\ \citenamefont
  {Weinberger}}]{Chuan2017OnCalibration}%
  \BibitemOpen
  \bibfield  {author} {\bibinfo {author} {\bibfnamefont {C.}~\bibnamefont
  {Guo}}, \bibinfo {author} {\bibfnamefont {G.}~\bibnamefont {Pleiss}},
  \bibinfo {author} {\bibfnamefont {Y.}~\bibnamefont {Sun}}, \ and\ \bibinfo
  {author} {\bibfnamefont {K.~Q.}\ \bibnamefont {Weinberger}},\ }in\ \href
  {https://proceedings.mlr.press/v70/guo17a.html} {\emph {\bibinfo {booktitle}
  {Proceedings of the 34th International Conference on Machine Learning}}},\
  \bibinfo {series} {Proceedings of Machine Learning Research}, Vol.~\bibinfo
  {volume} {70},\ \bibinfo {editor} {edited by\ \bibinfo {editor}
  {\bibfnamefont {D.}~\bibnamefont {Precup}}\ and\ \bibinfo {editor}
  {\bibfnamefont {Y.~W.}\ \bibnamefont {Teh}}}\ (\bibinfo  {publisher} {PMLR},\
  \bibinfo {year} {2017})\ pp.\ \bibinfo {pages} {1321--1330}\BibitemShut
  {NoStop}%
\bibitem [{Note1()}]{Note1}%
  \BibitemOpen
  \bibinfo {note} {Note, for a fixed Clifford circuit, the purification axis
  does not depend on $\protect \mathcal {M}_T$.}\BibitemShut {Stop}%
\bibitem [{\citenamefont {Hinton}\ and\ \citenamefont
  {Salakhutdinov}(2006)}]{hinton2006reducing}%
  \BibitemOpen
  \bibfield  {author} {\bibinfo {author} {\bibfnamefont {G.~E.}\ \bibnamefont
  {Hinton}}\ and\ \bibinfo {author} {\bibfnamefont {R.~R.}\ \bibnamefont
  {Salakhutdinov}},\ }\href@noop {} {\bibfield  {journal} {\bibinfo  {journal}
  {science}\ }\textbf {\bibinfo {volume} {313}},\ \bibinfo {pages} {504}
  (\bibinfo {year} {2006})}\BibitemShut {NoStop}%
\bibitem [{\citenamefont {LeCun}\ \emph {et~al.}(2015)\citenamefont {LeCun},
  \citenamefont {Bengio},\ and\ \citenamefont {Hinton}}]{lecun2015deep}%
  \BibitemOpen
  \bibfield  {author} {\bibinfo {author} {\bibfnamefont {Y.}~\bibnamefont
  {LeCun}}, \bibinfo {author} {\bibfnamefont {Y.}~\bibnamefont {Bengio}}, \
  and\ \bibinfo {author} {\bibfnamefont {G.}~\bibnamefont {Hinton}},\
  }\href@noop {} {\bibfield  {journal} {\bibinfo  {journal} {nature}\ }\textbf
  {\bibinfo {volume} {521}},\ \bibinfo {pages} {436} (\bibinfo {year}
  {2015})}\BibitemShut {NoStop}%
\bibitem [{\citenamefont {Goodfellow}\ \emph {et~al.}(2016)\citenamefont
  {Goodfellow}, \citenamefont {Bengio},\ and\ \citenamefont
  {Courville}}]{goodfellow2016deep}%
  \BibitemOpen
  \bibfield  {author} {\bibinfo {author} {\bibfnamefont {I.}~\bibnamefont
  {Goodfellow}}, \bibinfo {author} {\bibfnamefont {Y.}~\bibnamefont {Bengio}},
  \ and\ \bibinfo {author} {\bibfnamefont {A.}~\bibnamefont {Courville}},\
  }\href@noop {} {\emph {\bibinfo {title} {Deep learning}}}\ (\bibinfo
  {publisher} {MIT press},\ \bibinfo {year} {2016})\BibitemShut {NoStop}%
\bibitem [{\citenamefont {Lawrence}\ \emph {et~al.}(1997)\citenamefont
  {Lawrence}, \citenamefont {Giles}, \citenamefont {Tsoi},\ and\ \citenamefont
  {Back}}]{lawrence1997face}%
  \BibitemOpen
  \bibfield  {author} {\bibinfo {author} {\bibfnamefont {S.}~\bibnamefont
  {Lawrence}}, \bibinfo {author} {\bibfnamefont {C.~L.}\ \bibnamefont {Giles}},
  \bibinfo {author} {\bibfnamefont {A.~C.}\ \bibnamefont {Tsoi}}, \ and\
  \bibinfo {author} {\bibfnamefont {A.~D.}\ \bibnamefont {Back}},\ }\href@noop
  {} {\bibfield  {journal} {\bibinfo  {journal} {IEEE transactions on neural
  networks}\ }\textbf {\bibinfo {volume} {8}},\ \bibinfo {pages} {98} (\bibinfo
  {year} {1997})}\BibitemShut {NoStop}%
\bibitem [{\citenamefont {Bairey}\ \emph {et~al.}(2020)\citenamefont {Bairey},
  \citenamefont {Guo}, \citenamefont {Poletti}, \citenamefont {Lindner},\ and\
  \citenamefont {Arad}}]{Bairey2020Learning}%
  \BibitemOpen
  \bibfield  {author} {\bibinfo {author} {\bibfnamefont {E.}~\bibnamefont
  {Bairey}}, \bibinfo {author} {\bibfnamefont {C.}~\bibnamefont {Guo}},
  \bibinfo {author} {\bibfnamefont {D.}~\bibnamefont {Poletti}}, \bibinfo
  {author} {\bibfnamefont {N.~H.}\ \bibnamefont {Lindner}}, \ and\ \bibinfo
  {author} {\bibfnamefont {I.}~\bibnamefont {Arad}},\ }\href {\doibase
  10.1088/1367-2630/ab73cd} {\bibfield  {journal} {\bibinfo  {journal} {New
  Journal of Physics}\ }\textbf {\bibinfo {volume} {22}},\ \bibinfo {pages}
  {032001} (\bibinfo {year} {2020})}\BibitemShut {NoStop}%
\bibitem [{\citenamefont {James}\ \emph {et~al.}(2001)\citenamefont {James},
  \citenamefont {Kwiat}, \citenamefont {Munro},\ and\ \citenamefont
  {White}}]{James2001Measurement}%
  \BibitemOpen
  \bibfield  {author} {\bibinfo {author} {\bibfnamefont {D.~F.~V.}\
  \bibnamefont {James}}, \bibinfo {author} {\bibfnamefont {P.~G.}\ \bibnamefont
  {Kwiat}}, \bibinfo {author} {\bibfnamefont {W.~J.}\ \bibnamefont {Munro}}, \
  and\ \bibinfo {author} {\bibfnamefont {A.~G.}\ \bibnamefont {White}},\ }\href
  {\doibase 10.1103/PhysRevA.64.052312} {\bibfield  {journal} {\bibinfo
  {journal} {Phys. Rev. A}\ }\textbf {\bibinfo {volume} {64}},\ \bibinfo
  {pages} {052312} (\bibinfo {year} {2001})}\BibitemShut {NoStop}%
\bibitem [{\citenamefont {Barratt}\ \emph {et~al.}(2022)\citenamefont
  {Barratt}, \citenamefont {Agarwal}, \citenamefont {Potter}, \citenamefont
  {Gopalakrishnan},\ and\ \citenamefont {Vasseur}}]{barratt2022transitions}%
  \BibitemOpen
  \bibfield  {author} {\bibinfo {author} {\bibfnamefont {F.}~\bibnamefont
  {Barratt}}, \bibinfo {author} {\bibfnamefont {U.}~\bibnamefont {Agarwal}},
  \bibinfo {author} {\bibfnamefont {A.~C.}\ \bibnamefont {Potter}}, \bibinfo
  {author} {\bibfnamefont {S.}~\bibnamefont {Gopalakrishnan}}, \ and\ \bibinfo
  {author} {\bibfnamefont {R.}~\bibnamefont {Vasseur}},\ }\href@noop {}
  {\bibfield  {journal} {\bibinfo  {journal} {arXiv preprint arXiv:2206.12429}\
  } (\bibinfo {year} {2022})}\BibitemShut {NoStop}%
\bibitem [{\citenamefont {Brown}\ and\ \citenamefont {Fawzi}(2013)}]{Brown13}%
  \BibitemOpen
  \bibfield  {author} {\bibinfo {author} {\bibfnamefont {W.}~\bibnamefont
  {Brown}}\ and\ \bibinfo {author} {\bibfnamefont {O.}~\bibnamefont {Fawzi}},\
  }\href {\doibase 10.1109/isit.2013.6620245} {\bibfield  {journal} {\bibinfo
  {journal} {2013 IEEE International Symposium on Information Theory (ISIT)}\
  ,\ \bibinfo {pages} {346 }} (\bibinfo {year} {2013})}\BibitemShut {NoStop}%
\bibitem [{\citenamefont {Hastings}\ and\ \citenamefont
  {Haah}(2021)}]{Hastings21}%
  \BibitemOpen
  \bibfield  {author} {\bibinfo {author} {\bibfnamefont {M.~B.}\ \bibnamefont
  {Hastings}}\ and\ \bibinfo {author} {\bibfnamefont {J.}~\bibnamefont
  {Haah}},\ }\href {\doibase 10.22331/q-2021-10-19-564} {\bibfield  {journal}
  {\bibinfo  {journal} {Quantum}\ }\textbf {\bibinfo {volume} {5}},\ \bibinfo
  {pages} {564} (\bibinfo {year} {2021})}\BibitemShut {NoStop}%
\bibitem [{\citenamefont {Anshu}\ \emph {et~al.}(2021)\citenamefont {Anshu},
  \citenamefont {Arunachalam}, \citenamefont {Kuwahara},\ and\ \citenamefont
  {Soleimanifar}}]{Anshu21}%
  \BibitemOpen
  \bibfield  {author} {\bibinfo {author} {\bibfnamefont {A.}~\bibnamefont
  {Anshu}}, \bibinfo {author} {\bibfnamefont {S.}~\bibnamefont {Arunachalam}},
  \bibinfo {author} {\bibfnamefont {T.}~\bibnamefont {Kuwahara}}, \ and\
  \bibinfo {author} {\bibfnamefont {M.}~\bibnamefont {Soleimanifar}},\ }\href
  {\doibase 10.1038/s41567-021-01232-0} {\bibfield  {journal} {\bibinfo
  {journal} {Nature Phys.}\ }\textbf {\bibinfo {volume} {17}},\ \bibinfo
  {pages} {931} (\bibinfo {year} {2021})}\BibitemShut {NoStop}%
\bibitem [{\citenamefont {Haah}\ \emph {et~al.}(2021)\citenamefont {Haah},
  \citenamefont {Kothari},\ and\ \citenamefont {Tang}}]{Haah21}%
  \BibitemOpen
  \bibfield  {author} {\bibinfo {author} {\bibfnamefont {J.}~\bibnamefont
  {Haah}}, \bibinfo {author} {\bibfnamefont {R.}~\bibnamefont {Kothari}}, \
  and\ \bibinfo {author} {\bibfnamefont {E.}~\bibnamefont {Tang}},\ }\href@noop
  {} {\bibfield  {journal} {\bibinfo  {journal} {arXiv:2108.04842}\ } (\bibinfo
  {year} {2021})}\BibitemShut {NoStop}%
\bibitem [{\citenamefont {Huang}\ \emph {et~al.}(2020)\citenamefont {Huang},
  \citenamefont {Kueng},\ and\ \citenamefont {Preskill}}]{Huang2022Predicting}%
  \BibitemOpen
  \bibfield  {author} {\bibinfo {author} {\bibfnamefont {H.-Y.}\ \bibnamefont
  {Huang}}, \bibinfo {author} {\bibfnamefont {R.}~\bibnamefont {Kueng}}, \ and\
  \bibinfo {author} {\bibfnamefont {J.}~\bibnamefont {Preskill}},\ }\href
  {\doibase 10.1038/s41567-020-0932-7} {\bibfield  {journal} {\bibinfo
  {journal} {Nature Physics}\ }\textbf {\bibinfo {volume} {16}},\ \bibinfo
  {pages} {1050} (\bibinfo {year} {2020})}\BibitemShut {NoStop}%
\bibitem [{\citenamefont {Kuo}\ and\ \citenamefont
  {Dehghani}(2022)}]{Dehghani2022Unsupervised}%
  \BibitemOpen
  \bibfield  {author} {\bibinfo {author} {\bibfnamefont {E.-J.}\ \bibnamefont
  {Kuo}}\ and\ \bibinfo {author} {\bibfnamefont {H.}~\bibnamefont {Dehghani}},\
  }\href {\doibase 10.1103/PhysRevB.105.235136} {\bibfield  {journal} {\bibinfo
   {journal} {Phys. Rev. B}\ }\textbf {\bibinfo {volume} {105}},\ \bibinfo
  {pages} {235136} (\bibinfo {year} {2022})}\BibitemShut {NoStop}%
\bibitem [{\citenamefont {Towns}\ \emph {et~al.}(2014)\citenamefont {Towns},
  \citenamefont {Cockerill}, \citenamefont {Dahan}, \citenamefont {Foster},
  \citenamefont {Gaither}, \citenamefont {Grimshaw}, \citenamefont {Hazlewood},
  \citenamefont {Lathrop}, \citenamefont {Lifka}, \citenamefont {Peterson},
  \citenamefont {Roskies}, \citenamefont {Scott},\ and\ \citenamefont
  {Wilkins-Diehr}}]{xsede}%
  \BibitemOpen
  \bibfield  {author} {\bibinfo {author} {\bibfnamefont {J.}~\bibnamefont
  {Towns}}, \bibinfo {author} {\bibfnamefont {T.}~\bibnamefont {Cockerill}},
  \bibinfo {author} {\bibfnamefont {M.}~\bibnamefont {Dahan}}, \bibinfo
  {author} {\bibfnamefont {I.}~\bibnamefont {Foster}}, \bibinfo {author}
  {\bibfnamefont {K.}~\bibnamefont {Gaither}}, \bibinfo {author} {\bibfnamefont
  {A.}~\bibnamefont {Grimshaw}}, \bibinfo {author} {\bibfnamefont
  {V.}~\bibnamefont {Hazlewood}}, \bibinfo {author} {\bibfnamefont
  {S.}~\bibnamefont {Lathrop}}, \bibinfo {author} {\bibfnamefont
  {D.}~\bibnamefont {Lifka}}, \bibinfo {author} {\bibfnamefont {G.~D.}\
  \bibnamefont {Peterson}}, \bibinfo {author} {\bibfnamefont {R.}~\bibnamefont
  {Roskies}}, \bibinfo {author} {\bibfnamefont {J.~R.}\ \bibnamefont {Scott}},
  \ and\ \bibinfo {author} {\bibfnamefont {N.}~\bibnamefont {Wilkins-Diehr}},\
  }\href {\doibase 10.1109/MCSE.2014.80} {\bibfield  {journal} {\bibinfo
  {journal} {Computing in Science \& Engineering}\ }\textbf {\bibinfo {volume}
  {16}},\ \bibinfo {pages} {62} (\bibinfo {year} {2014})}\BibitemShut {NoStop}%
\bibitem [{\citenamefont {Nielsen}\ and\ \citenamefont
  {Chuang}(2002)}]{nielsen2002quantum}%
  \BibitemOpen
  \bibfield  {author} {\bibinfo {author} {\bibfnamefont {M.~A.}\ \bibnamefont
  {Nielsen}}\ and\ \bibinfo {author} {\bibfnamefont {I.}~\bibnamefont
  {Chuang}},\ }\href@noop {} {\enquote {\bibinfo {title} {Quantum computation
  and quantum information},}\ } (\bibinfo {year} {2002})\BibitemShut {NoStop}%
\bibitem [{\citenamefont
  {Gottesman}(1998{\natexlab{a}})}]{Gottesman1998Theory}%
  \BibitemOpen
  \bibfield  {author} {\bibinfo {author} {\bibfnamefont {D.}~\bibnamefont
  {Gottesman}},\ }\href {\doibase 10.1103/PhysRevA.57.127} {\bibfield
  {journal} {\bibinfo  {journal} {Phys. Rev. A}\ }\textbf {\bibinfo {volume}
  {57}},\ \bibinfo {pages} {127} (\bibinfo {year}
  {1998}{\natexlab{a}})}\BibitemShut {NoStop}%
\bibitem [{\citenamefont
  {Gottesman}(1998{\natexlab{b}})}]{gottesman1998heisenberg}%
  \BibitemOpen
  \bibfield  {author} {\bibinfo {author} {\bibfnamefont {D.}~\bibnamefont
  {Gottesman}},\ }\href@noop {} {\bibfield  {journal} {\bibinfo  {journal}
  {arXiv preprint quant-ph/9807006}\ } (\bibinfo {year}
  {1998}{\natexlab{b}})}\BibitemShut {NoStop}%
\bibitem [{\citenamefont {Aaronson}\ and\ \citenamefont
  {Gottesman}(2004)}]{aaronson2004Improved}%
  \BibitemOpen
  \bibfield  {author} {\bibinfo {author} {\bibfnamefont {S.}~\bibnamefont
  {Aaronson}}\ and\ \bibinfo {author} {\bibfnamefont {D.}~\bibnamefont
  {Gottesman}},\ }\href {\doibase 10.1103/PhysRevA.70.052328} {\bibfield
  {journal} {\bibinfo  {journal} {Phys. Rev. A}\ }\textbf {\bibinfo {volume}
  {70}},\ \bibinfo {pages} {052328} (\bibinfo {year} {2004})}\BibitemShut
  {NoStop}%
\bibitem [{\citenamefont {Cichosz}(2014)}]{cichosz2014data}%
  \BibitemOpen
  \bibfield  {author} {\bibinfo {author} {\bibfnamefont {P.}~\bibnamefont
  {Cichosz}},\ }\href@noop {} {\emph {\bibinfo {title} {Data mining algorithms:
  explained using R}}}\ (\bibinfo  {publisher} {John Wiley \& Sons},\ \bibinfo
  {year} {2014})\BibitemShut {NoStop}%
\end{thebibliography}
\end{document}